%% file: o3_frb.tex
\RequirePackage{lineno}
\documentclass[twocolumn,trackchanges]{aastex62}
\usepackage[utf8]{inputenc}

\usepackage{acronym,booktabs,multirow,cellspace,tabularx,xspace,mathtools}
\usepackage{breakurl}

\input{macros}

\begin{document}

\title{Search for Gravitational Waves Associated with Fast Radio Bursts Detected by CHIME/FRB During the LIGO--Virgo Observing Run O3a}

\input{combined_authors.tex}

\correspondingauthor{LIGO Scientific Collaboration, Virgo Collaboration, \& KAGRA Collaboration Spokespersons}
\email{lsc-spokesperson@ligo.org, virgo-spokesperson@ego-gw.it, hisaaki.shinkai@oit.ac.jp}
\date{\today}
\input{abstract}

\input{introduction}

\input{frb_gw_progenitors}
\input{chime_sample}

\input{search_methods}
\input{results}
\input{conclusions}
\acknowledgments
\input{acknowledgments}

\bibliographystyle{aasjournal}
\bibliography{bibliography}
\end{document}

%% file: macros.tex
\newcolumntype{Y}{>{\centering\arraybackslash}X}

\newacro{H1}{LIGO Hanford}
\newacro{L1}{LIGO Livingston}
\newacro{V1}{Virgo}
\newacro{FAST}{Five-hundred-meter Aperture Spherical radio Telescope}
\newacro{DQ shift}{Data Quality shift}
\newacro{FAP}{false alarm probability}
\newacro{EPO}{Education and Public Outreach}
\newacro{OURCA}{Office of Undergraduate Research and Creative Activity }
\newacro{DQ}{Data Quality}
\newacro{VLM}{Virginia Living Museum}
\newacro{AAPT}{American Association of Physics Teachers}
\newacro{SSS}{Second Saturday Stargazing}
\newacro{LIGO}{Laser Interferometer Gravitational Wave Observatory}
\newacro{aLIGO}{Advanced LIGO}
\newacro{AEI}{Albert Einstein Institute}
\newacro{CBC}{Compact Binary Coalescence}
\newacro{CDS}{Control and Data System}
\newacro{CIS}{Channel Information System}
\newacro{CIT}{Caltech}
\newacro{DAQ}{Data Acquisition System}
\newacro{DMP}{LIGO Data Management Plan}
\newacro{DMT}{Data Monitoring Toolbox}
\newacro{DQSEGDB}{Data Quality Segment Database}
\newacro{E@H}{Einstein at Home}
\newacro{ECSS}{Extended Collaborative Support Service}
\newacro{EM}{Electro-magnetic}
\newacro{ER}{Engineering Run}
\newacro{FTE}{Full-Time Equivalent}
\newacro{GCN}{Gamma-ray Coordinates Network}
\newacro{GLUE}{Grid LSC User Environment}
\newacro{GraCEDb}{Gravitational-wave Candidate Event Database}
\newacro{GRINCH}{Gravitational-wave Candidate Event Handlers}
\newacro{GSL}{GNU Scientific Library}
\newacro{GW}{gravitational wave}
\newacro{HTC}{High Throughput Computing}
\newacro{IFO}{Interferometer}
\newacro{ICTS}{International Centre for Theoretical Sciences}
\newacro{IdP}{Identity Provider}
\newacro{iLIGO}{Initial LIGO}
\newacro{IUCAA}{Inter-University Centre for Astronomy and Astrophysics}
\newacro{KDC}{Key Distribution Center}
\newacro{LALSuite}{LSC Algorithm Libraries}
\newacro{LIAM}{LIGO Identity and Access Management}
\newacro{LDR}{LIGO Data Replicator}
\newacro{LDG}{LIGO Data Grid}
\newacro{LIGOdv}{LIGO Data Viewer}
\newacro{LIGOdv-web}{LIGO Data Viewer Web Service}
\newacro{LOSC}{LIGO Open Science Center}
\newacro{LSC}{LIGO Scientific Collaboration}
\newacro{LSCSoft}{LSC Software Repositories}
\newacro{LSST}{Large Synoptic Survey Telescope}
\newacro{LVAlert}{LIGO-Virgo Alert System}
\newacro{LVC}{\ac{LSC} and the Virgo Collaboration}
\newacro{LVCN}{LIGO Virgo Computing Network}
\newacro{KAGRA}{Kamioka Gravitational Wave Detector}
\newacro{NDS}{Network Data Server}
\newacro{ODC}{Online Detector Characterization}
\newacro{OSG}{Open Science Grid}
\newacro{PE}{Parameter Estimation} 
\newacro{RDS}{Reduced Data Set}
\newacro{SDSC}{San Diego Supercomputer Center}
\newacro{SP}{Service Provider}
\newacro{TACC}{Texas Advanced Computing Center}
\newacro{XSEDE}{Extreme Science and Engineering Discovery Environment}
\newacro{VDT}{Virtual Data Toolkit}
\newacro{EO}{Engineering and Operations}
\newacro{COS}{Collaboration Operations Support}
\newacro{ABB}{Application Building Blocks}
\newacro{DHS}{Data Handling Services}
\newacro{DetChar}{Detector Characterization}
\newacro{GDS}{Global Diagnostics System}
\newacro{DA}{Data Analysis}
\newacro{SWIG}{the Simple Wrapper and Interface Generator}
\newacro{GRB}{gamma-ray burst}
\newacro{sGRB}[sGRB]{short gamma-ray burst}
\newacro{FRB}{fast radio burst}
\newacro{LIGO}{Laser Interferometer Gravitational-Wave Observatory}
\newacro{ASKAP}{Australian Square Kilometre Array Pathfinder}
\newacro{DetChar}{Detector Characterization}
\newacro{BBH}{binary black hole}
\newacro{EM}{electromagnetic}
\newacro{GCN}{Gamma-ray Coordination Network}
\newacro{Fermi}{Fermi Gamma-ray Burst Monitor}
\newacro{SWIFT}{The Neil Gehrels Swift Observatory}
\newacro{PCSE}{Physics, Computer Science and Engineering}
\newacro{CNU}{Christopher Newport University}
\newacro{IPN}{Interplanetary Gamma-Ray Burst Timing Network}
\newacro{MOU}{Memorandum of Understanding}
\newacro{NNPS}{Newport News Public Schools}

\newacro{CHIME}[CHIME]{the Canadian Hydrogen Intensity Mapping Experiment}
\newacro{RA}[RA]{right ascension}
\newacro{DEC}[Dec]{declination}
\newacro{ADI}[ADI]{accretion disk instability}
\newacro{BH}[BH]{black hole}
\newacro{BNS}[BNS]{binary neutron star \acused{NS}}
\newacro{CSG}[CSG]{circular sine--Gaussian}
\newacro{SG}[SG]{sine--Gaussian}
\newacro{GBM}[GBM]{Gamma-ray Burst Monitor}
\newacro{GCN}[GCN]{Gamma-ray Coordinates Network}
\newacro{FRB}[FRB]{fast radio burst}
\newacro{GW}[GW]{gravitational wave}
\newacro{NS}[NS]{neutron star}
\newacro{NSBH}[NSBH]{neutron star-black hole \acused{NS} \acused{BH}}
\newacro{O1}[O1]{the first observing run of Advanced LIGO and Advanced Virgo}
\newacro{O2}[O2]{the second observing run of Advanced LIGO and Advanced Virgo}
\newacro{O3}[O3]{the third observing run of Advanced LIGO and Advanced Virgo}
\newacro{O3a}[O3a]{the first part of the third observing run of Advanced LIGO and Advanced Virgo}
\newacro{O3b}[O3b]{the second part of the third observing run of Advanced LIGO and Advanced Virgo}
\newacro{O4}[O4]{the fourth observing run of Advanced LIGO, Advanced Virgo and Kagra}
\newacro{SNR}[SNR]{signal-to-noise ratio}
\newacro{VALID}[VALID]{Vetting Automation and Literature Informed Database}
\newacro{DM}{dispersion measure}

\newcommand{\LALSuite}{\texttt{LALSuite}\xspace}
\newcommand{\PYGRB}{\texttt{PyGRB}\xspace}
\newcommand{\PYCBC}{\texttt{PyCBC}\xspace}
\newcommand{\Xpipeline}{\texttt{X-Pipeline}\xspace}
\newcommand{\Msun}{\ensuremath{\textrm{M}_{\sun}}\xspace}
\newcommand{\dmu}{pc~cm\ensuremath{^{-3}}\xspace}

\newcommand{\OThreeAStart}{{1 April 2019 15:00 UTC}\xspace}
\newcommand{\OThreeAEnd}{{1 October 2019 15:00 UTC}\xspace}

\newcommand{\XpipeTotalAnalysesRepeatANDNonRepeat}{40\xspace} 
\newcommand{\XpipeTotalAnalysesNoRepeaters}{29\xspace}
\newcommand{\XpipeTotalAnalysesRepeaters}{11\xspace}
\newcommand{\TotalFRBsAnalysedNonRepeat}{34\xspace}

\newcommand{\pvalBurstLowest}{{\ensuremath{1.90\times 10^{-2}}\xspace}}  
\newcommand{\lowestRepeaterPValue}{{\ensuremath{1.3\times 10^{-1}}\xspace}} 
\newcommand{\EULSGA}{{\ensuremath{2.5~\times~10^{50}}~erg}\xspace}
\newcommand{\EULSGH}{{\ensuremath{7.9~\times~10^{54}}~erg}\xspace}
\newcommand{\EULRANGEWNB}{{\ensuremath{4.8-470~\times~10^{50}}~erg}\xspace}
\newcommand{\EULRANGEDSTWOP}{{\ensuremath{5.8-6.4~\times~10^{54}}~erg}\xspace}

\newcommand{\nCBC}{\ensuremath{22}\xspace}

\newcommand{\rDBNS}{\ensuremath{190}\xspace}

\newcommand{\rDNSBHGen}{\ensuremath{260}\xspace}

\newcommand{\rDNSBHAli}{\ensuremath{350}\xspace}
\newcommand{\pvalCBCLowestFAP}{\ensuremath{3.74\times 10^{-2}}\xspace}

%% file: combined_authors.tex
\author{R.~Abbott}
\affiliation{LIGO Laboratory, California Institute of Technology, Pasadena, CA 91125, USA}
\author{T.~D.~Abbott}
\affiliation{Louisiana State University, Baton Rouge, LA 70803, USA}
\author{F.~Acernese}
\affiliation{Dipartimento di Farmacia, Universit\`a di Salerno, I-84084 Fisciano, Salerno, Italy}
\affiliation{INFN, Sezione di Napoli, Complesso Universitario di Monte S. Angelo, I-80126 Napoli, Italy}
\author{K.~Ackley}
\affiliation{OzGrav, School of Physics \& Astronomy, Monash University, Clayton 3800, Victoria, Australia}
\author{C.~Adams}
\affiliation{LIGO Livingston Observatory, Livingston, LA 70754, USA}
\author{N.~Adhikari}
\affiliation{University of Wisconsin-Milwaukee, Milwaukee, WI 53201, USA}
\author{R.~X.~Adhikari}
\affiliation{LIGO Laboratory, California Institute of Technology, Pasadena, CA 91125, USA}
\author{V.~B.~Adya}
\affiliation{OzGrav, Australian National University, Canberra, Australian Capital Territory 0200, Australia}
\author{C.~Affeldt}
\affiliation{Max Planck Institute for Gravitational Physics (Albert Einstein Institute), D-30167 Hannover, Germany}
\affiliation{Leibniz Universit\"at Hannover, D-30167 Hannover, Germany}
\author{D.~Agarwal}
\affiliation{Inter-University Centre for Astronomy and Astrophysics, Pune 411007, India}
% ...

\collaboration{The LIGO Scientific Collaboration}
\collaboration{The Virgo Collaboration}
\collaboration{The KAGRA Collaboration}
\collaboration{The CHIME/FRB Collaboration}

%% file: abstract.tex
\begin{abstract}
We search for gravitational-wave transients associated with fast radio bursts (FRBs) detected by the Canadian Hydrogen Intensity Mapping Experiment Fast Radio Burst Project (CHIME/FRB), during the first part of the third observing run of Advanced LIGO and Advanced Virgo (1 April 2019 15:00 UTC\,--\,1 Oct 2019 15:00 UTC). Triggers from \nCBC FRBs were analyzed with a search that targets compact binary coalescences with at least one neutron star component. A targeted search for generic gravitational-wave transients was conducted on \XpipeTotalAnalysesRepeatANDNonRepeat FRBs.  We find no significant evidence for a gravitational-wave association in either search.  Given the large uncertainties in the distances of the FRBs inferred from the dispersion measures in our sample, however, this does not conclusively exclude any progenitor models that include emission of a gravitational wave of the types searched for from any of these FRB events.  We report $90\%$ confidence lower bounds on the distance to each FRB for a range of gravitational-wave progenitor models.  By combining the inferred maximum distance information for each FRB with the sensitivity of the gravitational-wave searches, we set upper limits on the energy emitted through gravitational waves for a range of emission scenarios. We find values of order $10^{51}$--$10^{57}$\,erg for a range of different emission models with central gravitational wave frequencies in the range 70--3560\,Hz. Finally, we also found no significant coincident detection of gravitational waves with the repeater, FRB\,20200120E, which is the closest known extragalactic FRB. 
\vspace{20mm}

\end{abstract}

%% file: introduction.tex
\section{Introduction}
\label{Introduction}

\indent Fast radio bursts (FRBs) are bright millisecond duration radio pulses that have been observed out to cosmological distances, several with inferred redshifts greater than unity \citep{Lorimer2007,Petroff:2019tty,CordesChatterjee2019ARA&A}. Although intensely studied for more than a decade, the emission mechanisms and progenitor populations of FRBs are still one of the outstanding questions in astronomy. \acused{FRB}

Some FRBs have been shown to repeat \citep{2nd_source_repeating, 8_new_repeating, Kumar_2019}, and the recent association of a FRB with the Galactic magnetar SGR 1935+2154 proves that magnetars can produce FRBs \citep{chime_sgr_2020Natur}. Alternative progenitors and mechanisms to produce non-repeating FRBs are still credible and have so far not been ruled out \citep{Zhang2020Natur}. Data currently suggests that both repeating and non-repeating classes of FRBs have Dispersion Measures (DMs), a quantity equal to the integral of the free electron density along the line of sight, and sky locations consistent with being drawn from the same population.  However, the two classes have been shown to differ in their intrinsic temporal widths and spectral bandwidths \citep{chime_cat_2021}. Whether genuine non-repeating sources have a different origin to their repeating cousins is an unresolved question. \acused{DM}

The first discovery of an FRB was made over a decade ago by Parkes 64m radio telescope \citep{Lorimer2007}. This burst, FRB~010724 or FRB~20010724A, known as the \emph{Lorimer burst}, first indicated an extragalactic origin for FRBs through its observed DM. This burst had a DM of 375~\dmu, far in excess of the likely Galactic DM contribution along the line of sight (of order 45~\dmu for this event), supporting an extragalactic origin. The precise localizations of FRB host galaxies have since unambiguously confirmed an extragalactic hypothesis \citep{2017Natur.541...58C,Bannister2019Sci,Li2020ApJ,Heintz_2020ApJ} and constraints on the progenitor population are starting to be understood \citep[e.g.][]{Bhandari2020ApJ}. The inferred cosmological distances for many FRBs have shown that these transients have extreme luminosities by radio standards, of the order $10^{38} - 10^{46}~\mathrm{erg}~\mathrm{s}^{-1}$ \citep{Zhang2018ApJ}.

Recent studies suggest a volumetric rate of order $3.5^{+5.7}_{-2.4} \times 10^{4}$ Gpc$^{-3}$yr$^{-1}$ above $10^{42}$\,erg\,s$^{-1}$ \citep{Luo_2020MNRAS}. Up to mid-2018, around 70 FRBs had been publicly announced \citep{Petroff2016PASA_catalogue}, of which around 7\% had been shown to repeat. The majority of the detections during this period had been made by Parkes \citep[27 FRBs at $\sim$ 1.5~GHz;][]{Champion2016MNRAS,Thornton2013Sci} and ASKAP \citep[28 FRBs at central frequencies of $\sim$ 1.3~GHz;][]{Bannister2017ApJ,shannonNature2018_dispersionbrightness}. Other detections were contributed by telescopes including UTMOST \citep{2017MNRAS_UTMOST} and the Green Bank Telescope \citep{Masui2015Natur_GBT}, each operating around 800~MHz, and  Arecibo \citep{2014ApJ_aricibo}, operating around $\sim$ 1.5~GHz. 

The FRB detection rate has greatly increased since \ac{CHIME} instrument \citep[][;see \url{https://chime-experiment.ca/}]{Newburgh2014SPIE,Bandura2014SPIE, chime_url} began its commissioning phase in late 2018, and its first FRB observation run shortly after. The CHIME radio telescope observes in the frequency range $400 - 800$~MHz and consists of four 20~m\,$\times$\,100~m cylindrical parabolical reflectors. Its large collecting area and wide field-of-view ($\approx 200$~deg${^2}$) make it a valuable survey instrument for radio transients. FRB detection for this instrument has been led by the CHIME/FRB project \citep{chime_frb_project_2018} which published its first sample of 13 FRBs during its early commissioning phase, despite operating at a lower sensitivity and field-of-view than design specifications \citep{Amiri2019Natur}.

The CHIME/FRB project recently published a catalog of 535 FRBs detected during their first year of operation; this includes 62 bursts from 18 previously identified repeating sources \citep{chime_cat_2021}. This is the first large collection of FRBs from a homogeneous survey and represents a significant milestone in this area of study. 
The CHIME/FRB data is supportive of different propagation or emission mechanisms between repeaters and non-repeaters, however, it is still not clear whether all FRBs do repeat \citep{Ravi2019NatAs} and, significantly, the FRB emission mechanism remains unknown. There presently exist many competing FRB emission theories \citep{Platts_2019}, some of which predict the accompaniment of a time-varying mass quadrupole moment, and thus, the emission of \acp{GW}. 

A number of studies have looked at the possibility of \ac{GW} emission associated with \acp{FRB} indirectly, using radio observations to search for coherent FRB-like emissions associated with short, hard \acp{GRB} \citep{Anderson2018ApJ,RowlinsonAnderson2019MNRAS,Gourdji_2020MNRAS,Rowlinson2020}. A radio search for FRB-like signals using early warning \ac{GW} alerts has also been suggested \citep{James2019MNRAS}.

The identification of an FRB within the sensitive reach of \ac{GW} interferometric detectors could provide conclusive proof of an association or constrain the parameters of the emission mechanisms for a given FRB. The increased population of detected FRBs from the CHIME/FRB Project therefore offers a unique chance of achieving this endeavor.

A first search for GW counterparts to transient radio sources was conducted by \citet{ligo_frb_search_2016}. This used a minimally modelled coherent search (\Xpipeline) $ \pm~2$~min around the detection time of 6 Parkes FRBs
using GW data from GEO600 \citep{Grote:2010zz} and initial Virgo \citep{VIRGO:2012dcp}. No GW coincidences were found, but this study provided a useful framework for future searches using improved GW sensitivities.  

In this paper we present the second targeted GW follow-up of FRBs using bursts detected by CHIME/FRB during \ac{O3a} \citep{aligo, avirgo}, which took place between \OThreeAStart and \OThreeAEnd.  This search uses both a generic GW transient search and a modelled search targeting coalescing binary systems. 

The organization of this paper is as follows: in Section \ref{sec:frb_gw_counterparts} we describe the motivation of this study by discussing possible GW counterparts to FRBs. We introduce the CHIME/FRB data sample in Section \ref{sec:frb_sample} and in Section \ref{sec:search_methods} discuss the GW search methods employed; this includes an overview of both of the pipelines used in our analysis. Section \ref{sec:results} provides the results of the GW analysis of the FRB sample. In Section \ref{sec:M81_burst} we report results of a gravitational wave analysis of the repeater, FRB\,20200120E, which is the closest known extragalactic FRB. Finally, in section \ref{sec:astrophysical-implications} we summarize the astrophysical implications of our results and discuss future GW searches for FRB counterparts at greater GW sensitivities.

%% file: frb_gw_progenitors.tex
\section{Proposed Gravitational Wave counterparts to FRBs}
\label{sec:frb_gw_counterparts}

This section will review some of the more popular models of non-repeating and repeating FRBs that could provide plausible GW counterparts and could therefore be constrained or confirmed through GW searches. (An online theory catalog tracks new FRB models; see \url{https://frbtheorycat.org}).

As the millisecond durations of FRBs indicate compact emission regions, many models of non-repeating FRBs have suggested cataclysmic events, including coalescing compact objects. 
A number of studies have investigated the possibility of FRB-like emissions from \ac{BNS} coalescence around the 
time of merger \citep[see review in][]{Platts_2019}. During this phase the magnetic fields of the \acp{NS} are synchronized to binary rotation and a coherent radiation could be generated due to magnetic braking. The mechanism requires magnetic fields of order $10^{12}$--$10^{13}$ Gauss and would produce FRB pulse widths consistent with the timescale of the orbital period of the \ac{BNS} just prior to coalescence \citep{totani_cosmological_2013}.

\citet{Wang2016ApJ} considered that an FRB could be produced during the final stages of a \ac{BNS} inspiral through magnetic reconnection due to the interaction of a toroidal magnetic field, produced as the \ac{NS} magnetospheres approach each other. Dynamic ejecta launched shortly after the final merger would produce significant opacity over a large solid angle, thus screening an FRB-type signal via absorption \citep{Yamasaki2018PASJ}. \citet{Zhang2020ApJ} has recently entertained the idea that similar interactions between the two \ac{NS} magnetospheres could produce repeating FRB-like coherent radio emissions decades or centuries before the final plunge.

Other studies have suggested that \ac{BNS} mergers could generate prompt coherent radio emission on ms timescales through mechanisms such as excitation of the circumbinary plasma by \ac{GW}s \citep{Moortgat_2005}, from a dynamically-generated magnetic field after the merger \citep{Pshirkov2010} or from the onset of the collision of a \ac{GRB} forward shock with the surrounding medium \citep{Usov_2000a,Sagiv_2002a}.

\citet{Zhang2016ApJ} postulated that the inspiral of a pair of spinning \acp{BH} could produce a Poynting flux, if at least one them is charged, by inducing a global magnetic dipole normal to the orbital plane (one of the black holes would require a characteristic charge of order $3.3 \times 10^{21}$ C (M/\Msun)). During the inspiral, as the orbital separation decreases, the magnetic flux of the system would change rapidly to produce particle bunching and thus, emission of coherent curvature radiation. The theory was extended in \citet{Zhang2019} to show that the methodology could also be applied to \ac{BNS} and \ac{NSBH} systems; it was termed the \emph{charged compact binary coalescence} signal. However, \citet{Zhang2019} showed that the relatively small charge sustained by the \acp{NS} would mean that the radio signal would be orders of magnitude dimmer than observed FRB events. Additionally, as in the case of \ac{BNS} mergers, the opacity from dynamic ejecta launched during the merger would negate an FRB-type signal. However, for systems with a mass ratio $m_{1}/m_{2} \gtrsim 5$ \citep{Shibata2009}, this process could produce an FRB as the \ac{NS} would plunge into the \ac{BH} with no tidal disruption.

Mergers of significant fractions of BNSs are likely to give rise to millisecond magnetars \citep{GaoZhang_2016PhRvD,MargalitBergerMetz2019ApJ}, although this is highly dependent on the unknown nuclear equation of state \citep[see][for a review]{SarinLasky_2021GReGr}. If the remnant \ac{NS} mass is greater than the maximum non-rotating
mass, it can survive for hundreds to thousands of seconds before collapsing to
form a BH \citep{Ravi2014MNRAS}. As the magnetic field lines snap as they cross the \ac{BH} horizon, an outwardly directed magnetic shock would dissipate as a short, intense radio burst \citep{Falcke_Rezzolla2014AA,Zhang_FRBsGRBs_2014ApJ}. This model has been motivated by the observation of relatively long lived X-ray plateaus following \acp{sGRB} that exhibit an abrupt decay phase, commonly interpreted as the collapse of the nascent \ac{NS} to a \ac{BH} \citep{Troja_2007,Lyons_2010,Rowlinson2010MNRAS, Rowlinson2013MNRAS}. Such collapses are expected to occur $\lesssim~5~\times~10^{4}$~s after the merger \citep{Ravi2014MNRAS}.

It has been suggested that FRBs could be related to the activity of magnetars or to strong pulses of energetic radio pulsars \citep{popov_millisecond_2013}. 
Additionally, the energy stored in rotational kinetic energy and the magnetic field of a millisecond pulsar are ample to power a repeating FRB \citep{Metzger2017ApJ}.  Resonant oscillation modes in the core and crust of magnetars have been suggested to cause quasi-periodic oscillations observed in the X-ray tails of giant flares. If the process by which these FRBs are created also excites non-radial modes in the magnetars, then \ac{GW}s could simultaneously be produced \citep[e.g.][]{Levin_2011MNRAS,QuitzowJames2017CQGra}. The detection of a repeating FRB-like event associated with the Galactic magnetar SGR 1935+2154 makes this a possible candidate for repeated GW emissions for repeating FRBs. 

The stellar oscillation mode that couples strongest to GW emission is the fundamental f-mode. The frequency of this mode depends on the equation of state, however analyzes of the tidal deformability of GW170817 are consistent with \ac{NS} f-mode frequencies typically being around 2~kHz \citep{2017PhRvL.119p1101A, Monitor:2017mdv, Wen_2019, Abbott:2018exr}. This is above the most sensitive frequency of the Advanced LIGO/Virgo observatories. While early theoretical studies indicated the GW amplitude could be large enough for f-mode oscillations from Galactic magnetar flares to be observable by Advanced LIGO/Virgo \citep{Ioka_2001MNRAS,CorsiOwen2011PhRvD}, more sophisticated analyzes give much more pessimistic predictions \citep{Levin_2011MNRAS,ZLK_2012PhRvD}. Other modes such as gravity modes (known as g-modes - here the restoring force is buoyancy) and r-modes (where the restoring force is the Coriolis force) emit at frequencies closer to the most sensitive range for Advanced LIGO/Virgo, however these modes couple more weakly to gravitational modes, and are therefore not likely to be detectable in association with an FRB.

%% file: chime_sample.tex
\section{The CHIME/FRB sample}
\label{sec:frb_sample}

The CHIME/FRB data sample provided for this analysis consists of 338 bursts observed within \ac{O3a}. Out of this sample, 168 bursts have been published in the first CHIME/FRB catalog \citep{chime_cat_2021}.  Within the sample of 338 bursts, only events overlapping with up-time of at least one of the three \ac{GW} observatories  were considered for analysis.  Within this sub-sample, the selection of bursts that were analyzed was based on the inferred distance to each burst.  This selection will be described at the end of this section, after the calculation of the inferred distance is described.  

The data for each FRB includes localization information, a topocentric arrival time and a measure of the total DM. For each burst, a Transient Name Server (TNS; see \url{https://www.wis-tns.org}) designation was also provided. The TNS naming convention takes the form `FRB YYYYMMDDLLL' with YYYY, MM and DD the year, month and day information in UTC and LLL a string from `A' to `Z', then from `aaa' to `zzz', indicating reporting order on any given day.

The arrival time at the \ac{CHIME} instrument's location (topocentric) at $400$~MHz was converted to a de-dispersed arrival time using the \ac{DM} value associated with each event. This time was used as the central event time around which each GW search was conducted. 

The localization information of each FRB is in the form of up to 5 disjoint error regions of varied morphology centered around the region with the highest SNR; each separate localization ``island'' has a central value and a 90\% confidence uncertainty region. The different approaches to these localization data adopted by the generic transient and modelled search pipelines will be described in Section \ref{sec:search_methods}.

To determine a measure of the luminosity distance of each FRB we employ the Macquart relation \citep{Macquart2020Natur}. This relation maps the redshift to the quantity $\mathrm{DM}_{\mathrm{IGM}}$, which is the DM contribution from extragalactic gas along the line of sight; this can be obtained after all other contributions are subtracted.  Taking into account all contributions to the total \ac{DM}, the quantity $\mathrm{DM}_{\mathrm{T}}$, a measure of redshift can therefore be determined by solving:
\begin{equation}\label{eq_dm}
\begin{split}
\MoveEqLeft
\mathrm{DM}_{\mathrm{T}}(z) = \mathrm{DM}_{\mathrm{MW}} + \mathrm{DM}_{\mathrm{halo}} + \mathrm{DM}_{\mathrm{IGM}}(z)\\
& + \mathrm{DM}_{\mathrm{host}}(z)/(1 + z) \,,
\end{split}
\end{equation}

\noindent where $\mathrm{DM}_{\mathrm{MW}}$ is the Milky Way contribution to the DM along the line of sight, $\mathrm{DM}_{\mathrm{halo}}$ is the contribution from the Milky Way halo and $\mathrm{DM}_{\mathrm{host}}$ the contribution from the host galaxy, which is corrected by the cosmic expansion factor.  The estimates of $z$ are then converted to a luminosity distance assuming a `flat-$\Lambda$' cosmology with the cosmological parameters $\Omega_{\mathrm
m}=0.31$, $\Omega_{\mathrm \Lambda}=0.69$ and
\mbox{$H_{0}=67.8$ km s$^{-1}$ Mpc$^{-1}$} \citep{Planck2015}.

To determine redshift values for each FRB we employ the Bayesian Markov-Chain Monte Carlo (MCMC) sampling framework described in \citep{Bhardwaj_2021ApJ} with a posterior distribution defined by:
\begin{equation}\label{eq_bayes}
 \mathcal{P}(\hat{\theta}\, |\,\mathrm{DM}_{\mathrm{T,O}}) =  \frac{ \mathcal{L}(\mathrm{DM}_{\mathrm{T,O}}\, |\, \hat{\theta}\,) \, \mathcal{\pi}(\hat{\theta})}{\mathcal{Z}}\,,
\end{equation}
\noindent where $\mathcal{L}(\mathrm{DM}_{\mathrm{T,O}}\, |\, \hat{\theta}\,)$ is the likelihood distribution of the observed quantity $\mathrm{DM}_{\mathrm{T,O}}$ given the parameters $\hat{\theta}$, $\mathcal{\pi}(\hat{\theta})$ are the prior distributions on $\hat{\theta}$ and $\mathcal{Z}$ is the Bayesian evidence; this latter factor enters Eq. (\ref{eq_bayes}) as a normalization factor independent of the model parameters and can be ignored if one is only interested in the posterior distribution rather than model selection. We assume a Gaussian likelihood function provided as:
\begin{equation}\label{loglike}
 \hspace{-1.5mm} \mathcal{L}(\mathrm{DM}_{\mathrm{T,O}}\, |\, \hat{\theta}\,) = \frac{1}{\sqrt{2 \pi \sigma^{2}} }
  \,\mathrm{exp}\left[  - \frac{(\mathrm{DM}_{\mathrm{T,O}} - \mathrm{DM}_\mathrm{T}(\hat{\theta} ))^{2}}{2\sigma^{2}} \right], 
\end{equation}
\noindent with $\sigma$ the uncertainty on $\mathrm{DM}_{\mathrm{T,O}}$ for each burst and $\mathrm{DM}_{\mathrm{T}}$ given by Eq. (\ref{eq_dm})  \citep{Rafiei-Ravandi:2021hbw}.

For the Milky Way contribution $\mathrm{DM}_{\mathrm{MW}}$, there is no consensus between the two popular models of \citet{NE2001_2002} and \citet{YMW16_2017ApJ}. Therefore, we follow \citet{Bhardwaj_2021ApJ} and assume a Gaussian prior based around the minimum of $\mathrm{DM}_{\mathrm{MW}}$ from these two models along the line of sight; a standard deviation of $20\%$ of this value is also used. 

The contribution $\mathrm{DM}_{\mathrm{halo}}$ has been estimated in a number of studies but is quite uncertain. For example, \citet{Yamasaki_2020ApJ} found values of $\mathrm{DM}_{\mathrm{halo}} \sim$ $30 - 245$~\dmu using a two component model. Studies by \citet{Dolag_2015MNRAS} found values between $\mathrm{DM}_{\mathrm{halo}} \sim~30 - 50$~\dmu based on cosmological simulation and \citet{Prochaska_2019MNRAS} estimated values between $30 - 80$~\dmu. To take account of the large uncertainty in this quantity we follow \citet{Bhardwaj_2021ApJ} and assume a Gaussian prior such that at $3\sigma$, $\mathrm{DM}_{\mathrm{halo}}$ has a value $0$ or $80$~\dmu.

The prior on $\mathrm{DM}_{\mathrm{IGM}}$ assumes the parameterization $\Delta = \mathrm{DM}_{\mathrm{IGM}} / \langle \mathrm{DM}_{\mathrm{IGM}}\rangle $ with the denominator obtained through the Macquart relation. This takes the form provided in \citet{Macquart2020Natur}:
\begin{equation}\label{eq_dm_prob}
P (\Delta) = A \Delta^{-\beta} \mathrm{exp} \left[ \frac{-(\Delta^{-\alpha} - C^{2})}{2 \alpha^2 \sigma_{\mathrm{DM}}^{2}} \right],
\end{equation}
\noindent with $\sigma_{\mathrm{DM}} = 0.2 z^{-0.5}$ and [$\alpha, \beta$] = 3; the value of $C$ is determined by requiring that $\langle \Delta \rangle = 1$. The form of this model is motivated by the requirement that the DM distribution approaches a Gaussian at small $\sigma_{\mathrm{DM}}$ in accordance with the Gaussianity of large scale structure. It also incorporates a skew at large $\sigma_{\mathrm{DM}}$ to reflect the possibility of over-densities along the line of sight. 

Finally, for a prior on $\mathrm{DM}_{\mathrm{host}}$, we adopt a lognormal distribution with median $e^{\mu}=68.2$ and logarithmic width parameter $\sigma_{\mathrm{host}}=0.88$ as in \citet{Macquart2020Natur}. 

The quantities outlined above have a large range of uncertainty and there could be additional contributions e.g., circumburst material. As a result, redshift values calculated from DMs are generally taken as upper limits. We perform MCMC sampling using the \texttt{emcee} package \citep{emcee2013PASP} based on an affine-invariant sampling algorithm \citep{GoodmanWeare2010} using 256 walkers of 20,000 samples. Inferred values of $z$, and thereby luminosity distance, and their 90\% credible intervals are thus determined for each FRB, based on the observed values of $\mathrm{DM}_{\mathrm{T}}$, \ac{RA} and \ac{DEC}, the estimated $\mathrm{DM}_{\mathrm{MW}}$ along the line of sight and the priors on other DM contributions described above. 

\begin{figure}[!t]
  \centering
  \includegraphics[width=\linewidth]{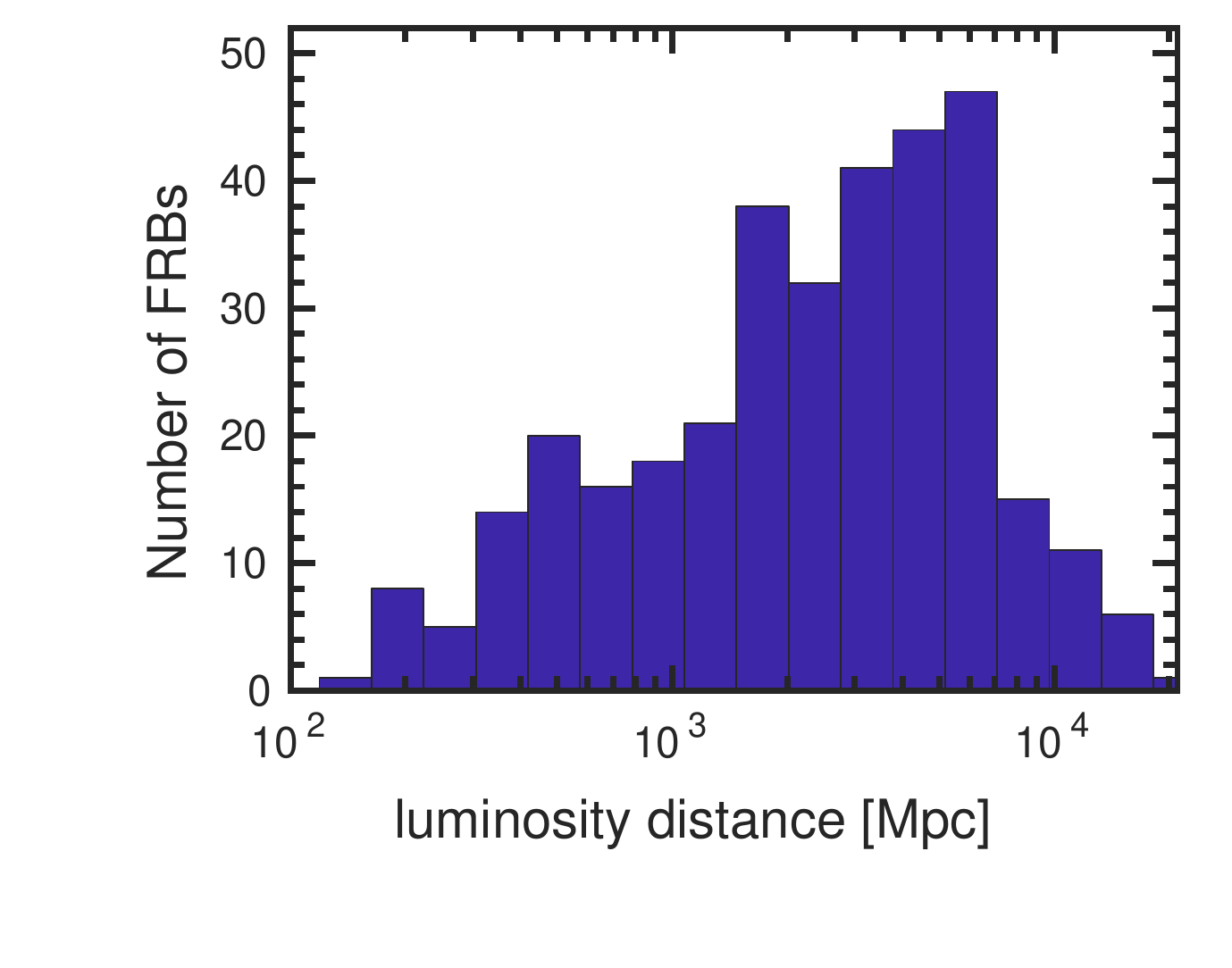}
    \caption{The distribution of inferred median distances for the CHIME/FRB data sample based on the MCMC analysis of Section \ref{sec:frb_sample}; there is a large uncertainty in these distances, thus this distribution should be taken as only an approximate representation. The distribution peaks between 1700~Mpc ($z \sim 0.3$) and 6000~Mpc ($z \sim 0.9$). The closest non-repeating event analyzed in our sample was FRB\,20190425A for which we inferred a median distance of 133~Mpc and a range [13--386]~Mpc at 90\% confidence; the most distant was FRB\,20190601C with a median inferred distance of 914 Mpc within a range [199--1737]~Mpc.
}
  \label{fig:frb_median_distance}
\end{figure}

Given the large uncertainties in the distances of FRBs, we based our analysis and results on the 90\% credible intervals inferred for the CHIME/FRB sample of bursts. However, for illustration, we show in Fig. \ref{fig:frb_median_distance} the distribution of the median distances of the total sample of 338 FRBs that occurred during \ac{O3a}. The plot shows that most events seem to occur within 1700~Mpc ($z \sim 0.3$) and 6000~Mpc ($z \sim 0.9$). The closest events in the distribution include a significant number of repeating FRBs. Due to the relatively limited range of the GW detectors, in selecting which bursts to analyze, we first downselected the sample to all bursts from the closest 10\% of CHIME/FRB non-repeating bursts that have GW detector network data available for analysis (if the recent CHIME/FRB catalog of 535 bursts is representative of the FRB population, at least around 11\% of FRBs repeat). Within this selection, a coherent analysis using modelled waveforms was then conducted on a smaller subset of the closest \nCBC non-repeating events for which data was available from at least one interferometric \ac{GW} detector, and a generic transient coherent analysis was conducted on a subset of FRBs for which data was available from at least two interferometric \ac{GW} detectors.  The further downselection to the final set of analyzes reported was based on two considerations.  For some events, the systematic noise in the detector was too significant near the time of the burst for one or both of our two searches, and these events were then excluded.  Finally, as each search requires significant personpower and computational resources, we performed searches on the remaining subset of events in order of increasing distance, until we reached a point of diminishing returns caused by the reduced overlap between the effective detection range of the GW detection network and the inferred distance to each FRB event. These considerations yielded a sample of \TotalFRBsAnalysedNonRepeat non-repeating FRBs that were analyzed by one or both types of analysis.  Using the same considerations for selection, we analyzed a total of \XpipeTotalAnalysesRepeaters repeated bursts from the closest 3 repeating sources: FRB\,20180916B (7 repeat events during O3A), FRB\,20180814A (2 repeat events) and FRB20190303A (2 events).  The lower and upper 90\% limits of the credible intervals on the luminosity distances to each of the non-repeating FRBs analyzed are included in the tables in Section \ref{sec:results}.

%% file: search_methods.tex
\section{Search Methods} 
\label{sec:search_methods}

Here we will provide a description of the two targeted search methods used in this paper.  These are the same methods applied to search for \ac{GW} events coincident with GRBs that occurred during the first \citep{2017ApJ...841...89A}, second \citep{2019ApJ...886...75A} and third \citep{2021ApJ...915...86A} Advanced LIGO and Advanced Virgo observing runs. In Section \ref{sec:cbc-search} we describe the modelled search method that aims to uncover sub-threshold \ac{GW} signals emitted by \ac{BNS} and \ac{NSBH} binaries \citep[\PYGRB;][]{Harry:2010fr, Williamson:2014wma}, highlighting choices in analysis configuration that are unique to the followup of FRB events.  In Section \ref{sec:Xpipeline-search} we discuss the search for generic \ac{GW} transients \citep[\Xpipeline;][]{Sutton2010NJPh, 2012PhRvD..86b2003W}.

\subsection{\PYGRB - Modelled search for binary mergers}
\label{sec:cbc-search}
The modelled search for \ac{GW}s associated with FRB events makes use of the \PYGRB data analysis pipeline \citep{Harry:2010fr, Williamson:2014wma}, and the search is configured to be similar to the search for \ac{GW} signals coincident with \acp{GRB} in \ac{O3a}  \citep{2021ApJ...915...86A}.  This is a coherent matched-filtering pipeline that compares the \ac{GW} detector network data with a bank of pre-generated waveforms, including the inspiral of \ac{BNS} and \ac{NSBH} binaries.  \PYGRB uses the \PYCBC \citep{alex_nitz_2020_3961510} open-source framework for distribution of the analysis of the \ac{GW} data across large computing clusters, and also relies on several elements of the \LALSuite software library \citep{LALSuite}.  

The \PYGRB analysis searches the combined detector data in the range $30$--$1000$~Hz.  A set of coherent data streams is formed by combining the data from the detectors, using a sample of sky-positions in the region reported for the FRB event that is being studied.  These data streams are then compared using matched filtering to the same predefined bank of waveform templates \citep{Owen:1998dk} used in the search for \acp{GW} associated with \acp{GRB} events in \ac{O3a}  \citep{2021ApJ...915...86A}.  The bank is created with a hybrid of geometric and stochastic template placement methods across target search space  \citep{Harry_2008, Brown_2012, Harry_2014, Capano:2016dsf, DalCanton:2017ala}, using a phenomenological inspiral-merger-ringdown waveform model for non-precessing point-particle binaries \citep[\texttt{IMRPhenomD};][]{Husa:2015iqa,  Khan:2015jqa}. This bank of templates is designed to cover binary masses in the range $[1.0, 2.8]\Msun$ for \acp{NS}, and $[1.0, 25.0]\Msun$ for \acp{BH}. The bank also allows for aligned-spin, zero-eccentricity \ac{BNS} and \ac{NSBH}, with dimensionless spins in the range $[0, 0.05]$ for \acp{NS} and
$[0, 0.998]$ for \acp{BH}. 

Coherent matched filtering can be susceptible to loud transient noise in the detector data and can produce a high \ac{SNR} \citep{2017ApJ...849..118N}. To combat this, the analysis performs additional tests on each point of high \ac{SNR} data, which we also refer to as triggers.  These tests can either remove the trigger or re-weight the \ac{SNR} using a $\chi^2$ test. This latter test determines how well the data agrees with the template over the whole template duration. Such cuts and re-weighting significantly improve the ability of the search to distinguish a GW from many types of transient noise, thus improving the significance of real GW triggers. The final re-weighted \ac{SNR} of each candidate event is used as the measure of its relative significance, or ranking statistic, within the search.

The \PYGRB analysis searches for \ac{GW} inspiral events that merge within 12~s of the de-dispersed event time of each FRB, with an asymmetric \textit{on-source window} starting 10~s before the FRB event and ending 2~s after the event. The search window is chosen to strike a balance between maximizing the possible progenitor models through a wider window or maximizing the sensitivity of the search by using a narrower window. In this search 
we seek a GW signal with a merger time close to the time of the FRB, assuming the FRB results from the interaction of the two binary components.  

The sensitivity of the search is governed by the comparison between the most significant event in the on-source window and the most significant event in equivalent trial searches of 12~s windows in the surrounding data, known as the \textit{off-source trials}. These off-source trials form the background data for the search, and if a sufficient number of background trials are conducted, this allows the search to determine the significance of any candidate events in the on-source window to the level needed to make a confident detection statement by computing a false-alarm probability. 

If multiple detectors are available, then additional effective background data can be produced by combining the data from the detectors with an intentional misalignment in time of at least the light-travel time across the network to ensure any detected events cannot possibly be true coherent GW candidates \citep{Williamson:2014wma}. This can be repeated for multiple possible time shifts, and in this search, these time shifts are set to match the on-source window length of 12~s.  This produces fewer time shifts than a 6~s on-source window, as used in previous searches for \ac{GW} associated with \ac{GRB} events such as \citet{2021ApJ...915...86A}.  This again impacts the effective significance of any detected events, because the amount of background data used by the search is limited by the amount of coherently analyzable data for all detectors in the network that surrounds the target time.  Thus, a search is only conducted if a minimum of 30~min of data are available.  

In the results section, we report the effective range of each search conducted as a 90\% exclusion distance, $D_{90}$.  This is calculated by first creating a set of simulated GW signals to inject into the off-source data, then attempting to find these injected signals with the standard search pipeline.  The signals are injected with amplitudes appropriate for a distribution of distances between their simulated origin and the detectors, and the $D_{90}$ distance is defined
as the distance within which 90\% of the injected simulated signals are recovered with a ranking statistic greater than the loudest on-source event.

Mirroring the approach taken in the \ac{O3a} search for GW events associated with GRB detections \citep{2021ApJ...915...86A}, the injected signals include \ac{BNS} systems with dimensionless spins in the range $-0.4$ to $0.4$, taken from observed pulsar spins \citep{Hessels:2006ze}, and are distributed uniformly in spin and with random orientations.  Injections also include aligned spin \ac{NSBH} binaries, and \ac{NSBH} binaries with generically oriented spins up to 0.98, motivated by X-ray binary observations \citep[e.g.,][]{Ozel:2010su, Kreidberg:2012ud, Miller:2014aaa}. The simulated signals are intentionally generated using different GW signal models than those used in the matched-filtering template bank, to approximate the target search space difference between the approximate templates used and the true GW signals.  In particular, the injected waveforms are identical to those used in the equivalent \ac{O3a} GRB event follow up analysis  \citep{2021ApJ...915...86A}.  Precessing \ac{BNS} signals
are simulated using the TaylorT2 time-domain, post-Newtonian inspiral
approximant \citep[\texttt{SpinTaylorT2};][]{Sathyaprakash:1991mt,
  Blanchet:1996pi, Bohe:2013cla, Arun:2008kb, Mikoczi:2005dn,
  Bohe:2015ana, Mishra:2016whh}, while \ac{NSBH} injected waveforms
are generated assuming a point-particle effective-one-body model tuned
to numerical simulations which can allow for precession effects from
misaligned spins \citep[\texttt{SEOBNRv3};][]{Pan:2013rra,
  Taracchini:2013rva, Babak:2016tgq}.  Again, identical to the injections used in \citet{2021ApJ...915...86A}, \ac{NS} masses for the injections are taken
between 1~\Msun and 3~\Msun from a normal distribution centered at
1.4~\Msun with a standard deviation of
0.2~\Msun  \citet{Kiziltan:2013oja} and 0.4~\Msun for \ac{BNS} and
\ac{NSBH} systems, respectively.  \ac{BH} masses are taken to be
between 3~\Msun and 25~\Msun from a normal distribution centered at
10~\Msun with a standard deviation of 6~\Msun.

Although this \PYGRB follow up of FRB events mirrors the search conducted for \acp{GW} associated with GRB events in \ac{O3a}  \citep{2021ApJ...915...86A} where appropriate, there were several differences in the choices of analysis parameters for the FRB analysis. The first major difference has been noted above, wherein a 12~s on-source window is used, which is double that of the GRB analysis.  This does reduce the significance of any detected signals, but has the benefit of allowing for more progenitor models where the EM emission occurs further in time from the peak of the GW emission.  

Another significant change was the method of determining the area of sky over which to search for the GW signals. The FRB data sample contains multiple localizations for each event, each with their own \ac{RA} and \ac{DEC} uncertainties.  This effectively creates multiple patches on the sky where the source could potentially reside. The effective GW network localization capability results in 90\% credible regions for detections on the order of $\approx10 - 10000$~deg$^2$, with an average of order $100$ deg$^2$.  In contrast, the multiple \ac{O3a} FRB sample localizations spanned only order $1$ deg$^2$ in total \citep{2020LRR....23}. The sensitivity of the search also did not vary significantly over the sky localizations, and so the final set of sky positions considered by the analysis was one circular patch on the sky with a size large enough to ensure coverage over all possible provided FRB localizations.  Within this patch, the sky is sampled by creating a circular grid of sky positions such that the time-delay between grid points is kept below $0.5$\,s \citep{Williamson:2014wma}.  This ensures coverage of the possible sky location of the source.  For each sky position, the timestream data from each GW detector is combined with the appropriately different time offsets required to form a coherent streams of data for that point on the grid.  These multiple coherent time streams are finally each considered in the search.

\subsection{\Xpipeline - Unmodelled search for generic transients}
\label{sec:Xpipeline-search}

The search for generic transients is performed with the coherent analysis algorithm \Xpipeline \citep{Sutton2010NJPh, 2012PhRvD..86b2003W}. This targeted search uses the sky localization and time window for each CHIME/FRB trigger to identify consistent excess power that is coherent across the network of GW detectors. We use different search parameters in our searches for repeating and non-repeating FRB sources.

There are a number of differences between our generic transient search on non-repeated sources and those previously conducted on GRBs \citep{2017ApJ...841...89A, 2019ApJ...886...75A, 2021ApJ...915...86A}. As in GRB searches, the on-source time window is chosen to start 600~s before the trigger, but is extended from 60~s seconds post trigger to 120~s to allow for the possibility of \ac{GW} emissions delayed relative to the FRB emission. This on-source window is also longer than the $\pm 120$~s window employed in the previous FRB search \citep {ligo_frb_search_2016}. The extended window allows for a greater number of non-\ac{CBC} sources than those considered in GRB searches and possible \ac{GW} emissions from magnetars, given the recent FRB-magnetar association \citep {chime_sgr_2020Natur}. 

The broadband search for FRBs with \Xpipeline covers the range 32~Hz up to 2~kHz, the upper range being higher than the GRB search (20--500~Hz) in order to include \ac{GW} emissions from oscillation modes of \ac{NS}s that are likely to occur above 1~kHz, specifically f-modes \citep{Wen_2019,nsfmodes}. 
We note that above 300~Hz a $\propto f^{2}$ frequency dependence in energy (see later Eq. (5)) combined with the $\propto f^{1}$ of the noise power spectral density of the detector increases the GW energy required to enable a confident detection as $\propto f^{3}$. Although including high frequency data increases the computational cost, including this data allows us to set limits on a wider variety of signal models.

\Xpipeline processes the on-source data around each FRB trigger by combining the \ac{GW} data coherently, taking into account the antenna response and noise level of each detector to generate a series of time--frequency maps. The maps show the temporal evolution of the spectral properties of the signal and allow searches for clusters of pixels with excess energy significantly greater than one would expect from background noise. These clusters are referred to as \textit{events}.

Events are given a ranking statistic based on energy and are subjected to coherent consistency tests based on the signal correlations between data in different detectors. This allows \Xpipeline to veto events that have properties similar to the noise background. 

The surviving event with the largest ranking statistic is taken to be the best candidate for a GW detection. Its significance is quantified as the probability for the background alone to produce such an event. This is done by comparing the \ac{SNR} of the trigger within the 720~s on-source to the distribution of the \acp{SNR} of the loudest triggers in the off-source trials. The off-source data are set to consist of at least $1.5$ hours of coincident data from at least two detectors around the trigger time.  This window is small enough to select data where the detectors should be in a similar state of operation as during the on-source interval, and large enough so that through artificial time-shifting, probabilities can be estimated at the sub-percent level.

We quantify the sensitivity of the generic transient search by injecting simulated signals into off-source data and recovering them. We account for calibration errors by jittering the amplitude and arrival time of the injections according to a Gaussian distribution representative of the typical calibration uncertainties expected in \ac{O3a}. We compute the percentage of injections that have a significance higher than the best event candidate and determine the amplitude at which this percentage is above 90\%; this value sets the upper limit.

As discussed in Section \ref{sec:frb_sample}, localization information for each FRB is in the form of up to 5 non-contiguous or overlapping error regions of varied morphology. Occasionally these islands can be dominated by the uncertainty of a single island. The sky position errors can span a few degrees or more in RA. This could result in a temporal shift causing a \ac{GW} signal to be rejected by a coherent consistency test \citep{2012PhRvD..86b2003W}. 
For each island we set up a circular grid around the central location of the island, with overlapping grid points discarded. A coherent data stream is formed from the GW detector data with an appropriate time offset for each point on the grid.  These data streams are then analyzed. Grid positions are large enough to cover the error radius and dense enough to ensure a maximum timing delay error, set as $1.25 \times 10^{-4}$\,s, is within 25\% of the signal period at our frequency upper limit of 2000~Hz. This is 4 times finer than GRB searches that typically analyze data up to a frequency cutoff of 500~Hz. Using this grid approach, the antenna responses change only slightly over sky position; of order a few percent over a few degrees \citep{GWGRB20142014PhRvD}. The responses are known to change rapidly near a null of the response; in such a case they are already negligible. 

A particular difference between this search and other searches focused on GRBs is the increased number of simulated waveform types used in this study. Given the uncertainty in plausible \ac{GW} emissions, we consider a larger range of generic burst scenarios, using an extended set of those used in both GRB and magnetar searches \citep{2021ApJ...915...86A,2019ApJ...874..163A}. Also, as we have no knowledge on whether or not FRBs are beamed along the rotation axis of the progenitor, all of our signal models correspond to elliptical and random polarization.

The waveforms chosen to cover the search parameter space are from 3 families that have different morphological characteristics: binary signals, generic burst-like signals and accretion disk instability (ADI) models. \Xpipeline is equally adept at detecting signals whose frequency decreases with time (ADI) and signals whose frequency increases with time (CBC models; \citet{scienceruns,2017ApJ...841...89A}). This paper reports the results for CBCs when obtained using the dedicated modelled search (described in Section \ref{sec:cbc-search}), so we will limit our discussions here to only the latter two waveform families.

\begin{table}
  \caption{The main parameters of the waveform injections used for the generic transient search.
  Models and their parameters have been chosen to cover as large a parameter space as possible. For all models the central frequencies are shown. We note that WNB models are defined by an additional frequency bandwidth, this parameter is shown in parenthesis. For the SG and WNB waveforms the duration parameter scales the width of the Gaussian envelope; for the DS2P models this parameter defines the decay time constant.  An asterisk (*) denotes waveforms used in the repeaters search only; $^c$ denotes waveforms with a circular polarization.}
  \begin{center}
\centering
\label{tab:table_frb_burst_injections}
\begin{tabular}{lcc}
\hline
\hline
Label &   Frequency          & Duration Parameter \\
      &    [Hz]             &    [ms]      \\
      \\
      \hline
      \multicolumn{3}{c}{Sine--Gaussian Chirplets}\\
      \hline
SG-A &         70              &   14       \\
SG-B &         90              &   11       \\
SG-C &         145             &  6.9       \\
SG-D &         290             &  3.4       \\
SG-E &         650             &  1.5       \\
SG-F &         1100            &  0.9       \\
SG-G &        1600            &  0.6       \\
SG-H &         1995            &  0.5       \\
SG-I* &         2600            &  0.38       \\
SG-J* &         3100            &  0.32       \\
SG-K* &         3560            &  0.28       \\
SG-L*$^c$ &         1600            &  0.6       \\
SG-M*$^c$ &         1995            &  0.5       \\
      \hline
            \multicolumn{3}{c}{Ringdowns}\\
                  \hline
DS2P-A&         1500            &  100      \\
DS2P-B&         1500            &  200      \\
      \hline
                  \multicolumn{3}{c}{White noise bursts}\\
      \hline
WNB-A &        150  (100--200)    &  11       \\
WNB-B &         150  (100--200)    &  100       \\
WNB-C &         550  (100--1000)    &  11       \\
WNB-D &         550  (100--1000)    &  100       \\
\hline
\end{tabular}
\end{center}
\end{table}

The generic burst-type waveforms are described in Table \ref{tab:table_frb_burst_injections}, where we list the most important parameters \citep[see also][]{2019PhRvD.100b4017A}. In all cases, to determine exclusion distances for this model family, we assume an optimistic emission of energy in GWs of $E_{\mathrm{GW}} = 10^{-2} \Msun \textrm{c}^{2}$ \citep{2021ApJ...915...86A}. Waveforms in this family aim to capture the general characteristics of a burst of \ac{GW} energy: 

\begin{description}
    \item[Sine--Gaussian] \acused{SG} These signals have been used previously to represent the GWs from stellar collapses. The models are defined in Eq. (1) of \citet{2017ApJ...841...89A} with a $Q$ factor of 9 and varying central frequency as shown in Table \ref{tab:table_frb_burst_injections}. They can also model f-modes in the core of a canonical \ac{NS}.  We therefore also include them in the search over repeating sources, and include \ac{SG} waveforms at additional frequencies listed in Table \ref{tab:table_frb_burst_injections}.  In order to better constrain some models, we also include circularly polarized \ac{SG} chirplets at the frequencies nearest the f-mode range (1600~Hz and 1995~Hz) in the search over repeated sources.
	 
	    \item[Ringdowns (DS2P)] These signals capture the form of damped sinusoids (DS2P) at a frequency of 1500~Hz and decay constants of 100~ms and 200~ms. 
	 
  \item[White Noise Bursts (WNB)] These signals mimic broad bursts of uncorrelated white noise, time-shaped by a Gaussian envelope. We use two models band-limited within frequencies of 100–200~Hz and 100–1000~Hz, and with time constants of 11~ms and 100~ms.
  
\end{description}

Following the predictions from oscillation modes for \ac{NS} starquakes \citep{Wen_2019,2019EPJA...55..117L}, the first two waveforms in this family (SG and DS2P) have been used in the search for \ac{GW}s associated with magnetar bursts \citep{2019ApJ...874..163A}.

We also consider a range of Accretion Disk Instability (ADI) models. These are long-lasting waveforms which are modelled to represent the \ac{GW} emissions from instabilities in a magnetically suspended torus around a rapidly spinning \ac{BH}. The model specifics and parameters used to generate the five types of ADI signals, designated ADI-A to ADI-E, are the same used in the previous searches \citep[see Table 1 of][]{2017ApJ...841...89A}.

The version of \Xpipeline used in this analysis has a new feature named autogating. This feature increases the sensitivity of the longer-duration ($\gtrsim10$~s) signals, previously limited by loud background noise transients \citep{2021ApJ...915...86A}. This technique gates the whitened data from a single detector if the average energy over a 1-second window exceeds a user-specified threshold. To minimize the possibility of a loud GW transient be gated, this procedure is canceled if the average energy at the same time in any other detector exceeds the threshold. 
 
\subsubsection{X-pipeline Search on Repeating FRBs}
A subset of \XpipeTotalAnalysesRepeaters of the FRBs that we analyze have been identified to repeat. Repeating FRBs are likely caused by a process distinct from those that produce singular FRBs; most notably they are unlikely to be associated with \ac{CBC} events.  We therefore only run the \Xpipeline generic transient search on these events, and we choose the parameters to provide maximal sensitivity to the GW transients that would most probably be produced by flaring magnetars. 

This search is similar to that for GW events associated with magnetars during \ac{O3} \citep{magnetar_2021}.  The frequency band of the search ranges from $50$~Hz to $4000$~Hz, which encapsulates the \ac{NS} f-mode frequency band, but excludes the lowest frequencies where nonstationary noise could potentially `pollute' the search statistics.  The search spans 8~s of time centered within one second of the arrival time of the FRB to ensure optimal sensitivity at the event time. Injected waveforms are chosen to reasonably model the f-modes of a canonical \ac{NS} as described in \citet{Kokkotas_2001}. This includes a series of \ac{SG} chirplets with a $Q$ factor of 9 and varying center frequencies as shown in Table \ref{tab:table_frb_burst_injections}.  We also neglect to use the autogating algorithm for noise transients as described above, as its tendency is also to gate fast injections such as \ac{SG}.  We also inject white noise bursts to estimate sensitivity at broadband frequency ranges. 

\subsection{RAVEN Coincident Analysis}
\label{sec:raven_method}
To perform a wider sweep of the \ac{O3a} data, we also looked for coincidences between these CHIME/FRB events and existing GW candidates using the tools of the Rapid, on-source VOEvent Coincidence Monitor \citep[RAVEN;][]{urban2016, cho2019low} to query the Gravitational-Wave Candidate Event Database GraceDB \citep{gracedb_cite}. This query to GraceDB tests whether any GW candidates were found by any of the modelled or generic transient low-latency GW search pipelines within a time window around the FRB events. The queries used the same on-source search windows as our modelled and generic transient searches, with [$-10$~s,$+2$~s] and [$-600$~s,$+120$~s] windows around the FRB triggers, respectively.  We then computed the joint false-alarm rate of any coincident GW candidate within these windows using the overall rate of FRB events in the CHIME/FRB sample calculated across the full span of the \ac{O3a} observing run and the false-alarm rate of the GW candidate.  

%% file: results.tex
\section{Results of Analysis}
\label{sec:results}

\subsection{Analysis Subsample}
We performed two different searches: for non-repeating FRBs, a \PYGRB modelled search was completed on a total of \nCBC FRB events and an \Xpipeline search for generic transient signals was completed on a total of \XpipeTotalAnalysesNoRepeaters non-repeaters and \XpipeTotalAnalysesRepeaters repeating FRBs.

\subsection{The false-alarm probability ($p$-value) distribution} 

The searches conducted for \ac{GW} counterparts returned no likely \ac{GW} signals in association with any of the analyzed repeating or non-repeating FRB events.  

The most significant events found by the \PYGRB search and the \Xpipeline search had $p$-values of \pvalCBCLowestFAP and \pvalBurstLowest, respectively. For the \Xpipeline analysis of the repeating FRBs, the lowest $p$-value was \lowestRepeaterPValue, corresponding to the repeat FRB\,20190702B of burst FRB\,20190303A, for which we analyzed 2 burst events.

\begin{figure}[!t]
  \centering
  \includegraphics[width=\linewidth]{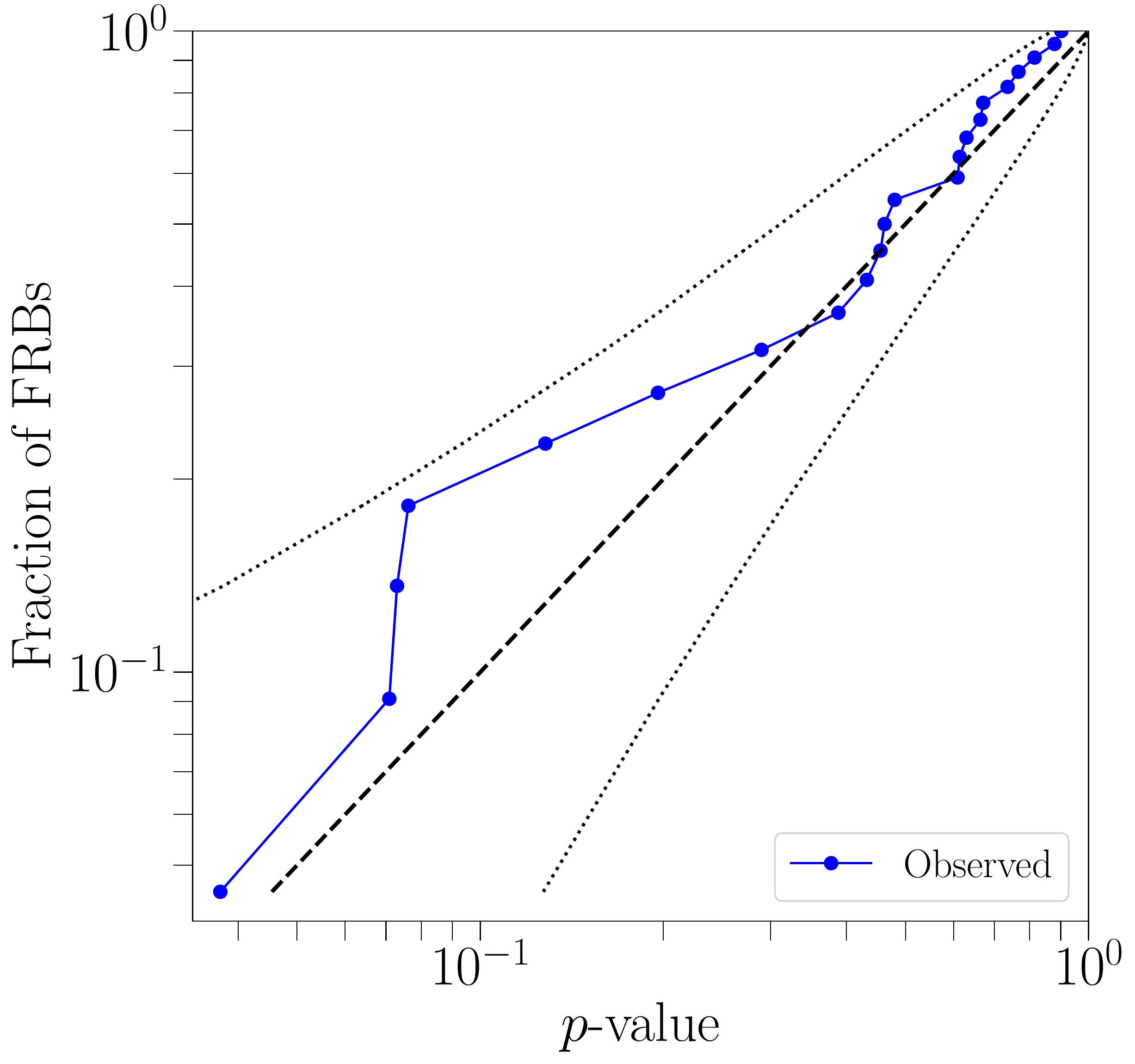}
  \caption{The cumulative distribution of $p$-values for the loudest on-source 
    events for the modelled search in \ac{O3a} around CHIME/FRB data. 
    The dashed line indicates an expected uniform 
    distribution of $p$-values under a no-signal hypothesis, with the 
    corresponding $90\%$ confidence band shown by the dotted lines.
}
  \label{fig:pygrb-pvalue}
\end{figure}

\begin{figure}[!t]
  \centering
  \includegraphics[width=\linewidth]{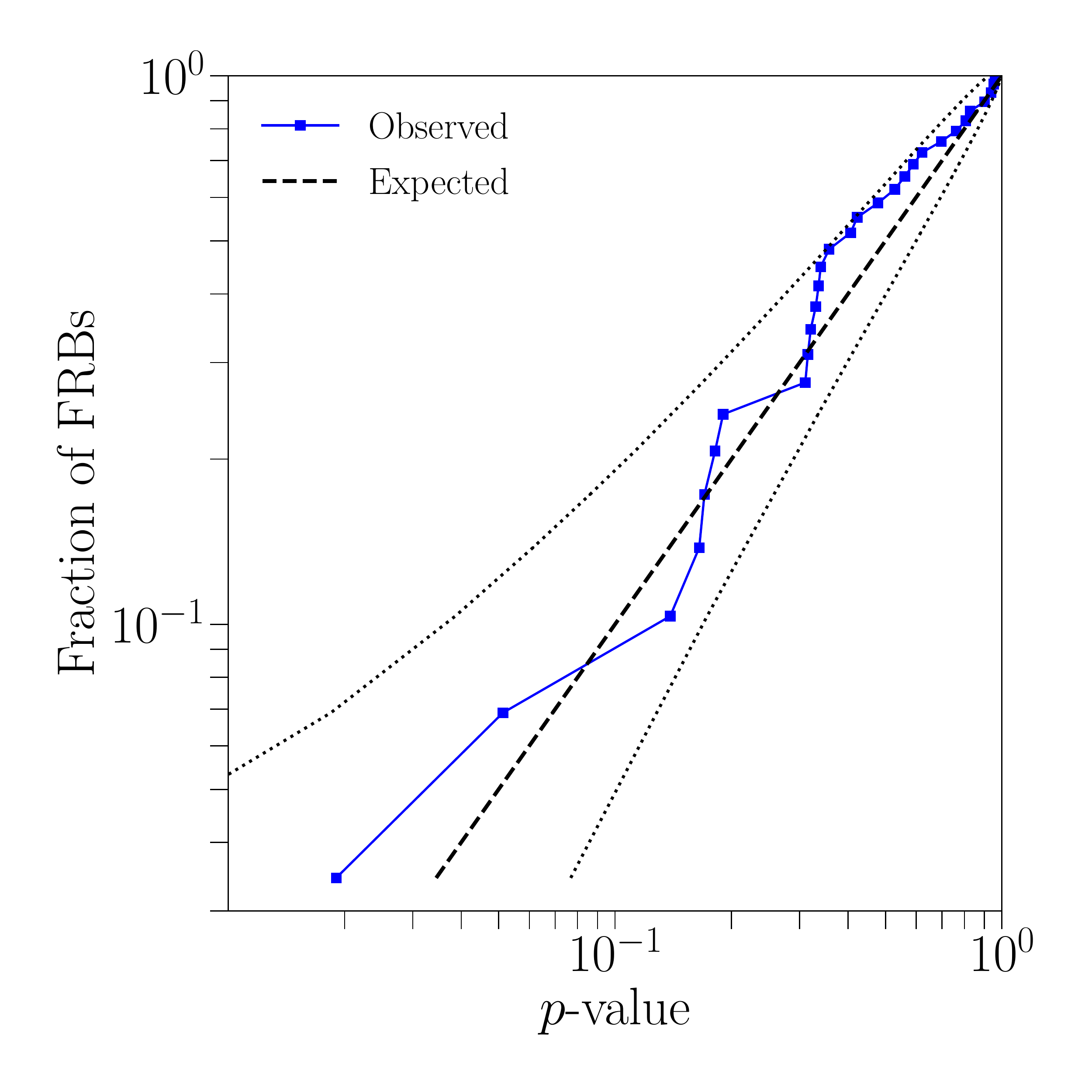}
  \caption{The cumulative distribution of $p$-values
    for the loudest events from the generic transient search for transient \acp{GW} associated with \XpipeTotalAnalysesNoRepeaters non-repeating CHIME/FRB bursts. The dashed line represents the expected
    distribution under the no-signal hypothesis, with the $90\%$ bands shown as dotted lines.
}
  \label{fig:burst-pvalue}
\end{figure}

The cumulative $p$-value distributions from both search methods are shown in Fig. \ref{fig:pygrb-pvalue} and Fig. \ref{fig:burst-pvalue}. In both figures, the dashed lines indicate the expected background distribution under the no-signal hypothesis, and the dotted lines indicate the 90\% confidence band around the no-signal hypothesis. 

\begin{figure}[]
  \centering
  \includegraphics[width=\linewidth]{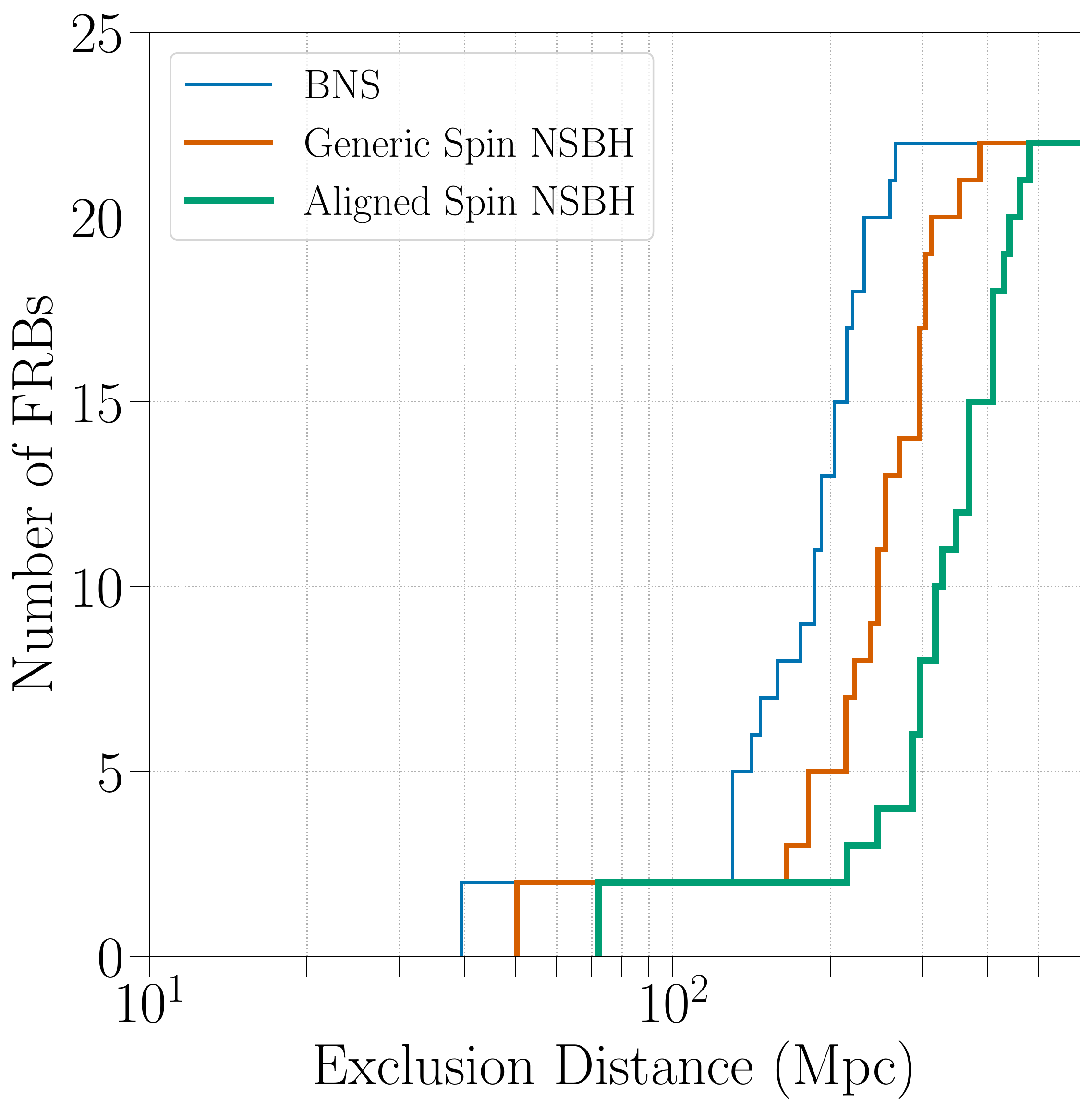}
  \caption{Cumulative histograms of the 90\% confidence exclusion
    distances, $D_{90}$, for the \nCBC CHIME/FRB bursts followed up by the modelled search. The blue line shows generically 
    spinning \ac{BNS} models, the orange line shows generically
    spinning \ac{NSBH} models, and the thick green line shows aligned spin \ac{NSBH} models. We define $D_{90}$ as the distance within which 90\% of the simulated \ac{GW}
    signals injected into the off-source data were recovered with a significance greater than the most significant on-source trigger.
}
  \label{pygrb-90exclusion}
\end{figure}

\begin{figure}[]
  \centering
  \includegraphics[width=\linewidth]{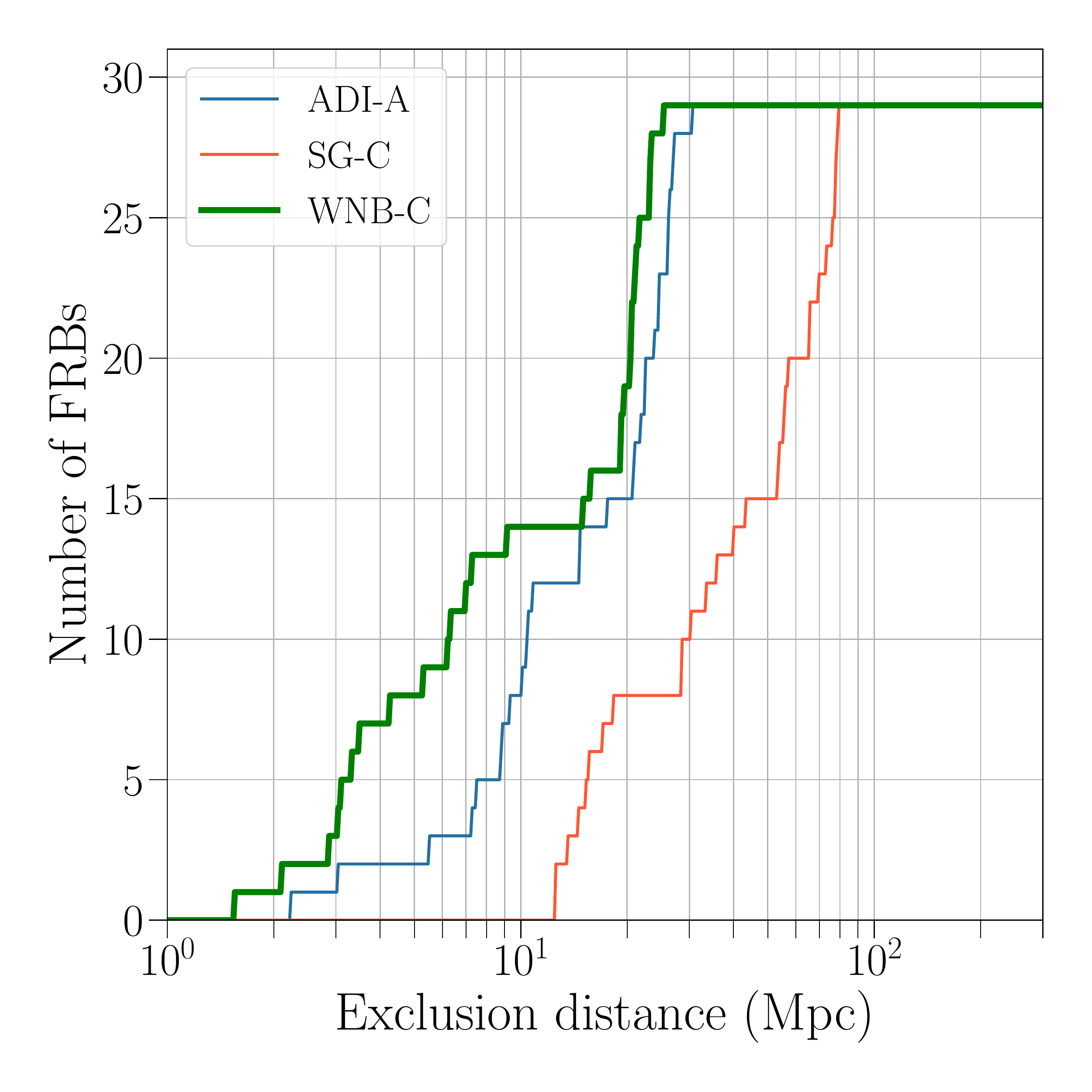}
  \caption{Cumulative histograms of the 90\%
    confidence exclusion distances, $D_{90}$, for  \ac{SG} model C (orange line), accretion disk instability
    (\ac{ADI}) signal model A (blue line) and white noise burst (WNB) model C (green, thick line). The quantity has the same definition as described in Fig. \ref{pygrb-90exclusion}. 
}
  \label{fig:burst-90exclusion}
\end{figure}
\subsection{Exclusion Distance Results}

Fig. \ref{pygrb-90exclusion} shows the cumulative $90\%$ exclusion distances for the \nCBC FRBs followed up with the modelled search. The lowest exclusion distances, of order 40~Mpc, were obtained for FRBs that occurred during times in which only Virgo data was available. 

For each of the three simulated signal classes considered in the modelled search, we quote the median of the $D_{90}$ results in the top row of Table \ref{tab:table_frb_median_exclusion_distance}; we see values of the order of $\rDBNS$~Mpc for \ac{BNS} and around \rDNSBHGen~Mpc (\rDNSBHAli~Mpc) for \ac{NSBH} with generic (aligned) spins. 

Fig. \ref{fig:burst-90exclusion} provides the cumulative $90\%$ exclusion distances for \XpipeTotalAnalysesNoRepeaters non-repeating FRBs considered in the generic transient search. This plot shows three representative burst models; ADI-A, SG-C and a WNB-C; the latter two have central frequencies of 145~Hz and 550~Hz respectively. Based on a standard $E_{\mathrm{GW}} \sim 10^{-2} \Msun \textrm{c}^{2}$ of emitted GW energy, there is a noticeable offset between the SG and the other two GW burst models. For the ADI-A waveform model, this is due to the energy of the former being distributed over a longer signal duration, of order $\sim40$~s; for the WNB-C model, this effect is due to a significant portion of its energy content being at higher frequency where detector performance is more comparatively limited.

\begin{table}
\setlength{\tabcolsep}{3pt} 
\caption{\label{tab:table_frb_median_exclusion_distance} Median values for the 90\% confidence level exclusion distances, $D_{90}$. Modelled search results are shown for three classes of \ac{BNS} progenitor model, and generic transient search results are shown for models described in Table \ref{tab:table_frb_burst_injections}.}
\begin{center}
 \begin{tabularx}{\columnwidth}{ c c c c }
\hline 
\hline 
\rule{0pt}{4ex} 
 Modelled &     & NSBH          & NSBH\\
  search    & BNS & Generic Spins & Aligned Spins\\ 
\hline 
\rule[-2ex]{0pt}{4ex} 
$D_{90}$ [Mpc] &  
  191.9 &  256.6 &  345.1 \\ 
\hline
\end{tabularx}
 \begin{tabularx}{\columnwidth}{  c X X X X }
\hline 
\rule{0pt}{4ex} 
 Unmodelled &    SG    & SG     & SG     & SG\\ 
  search   & A & B & C & D\\ 
\hline 
\rule[-2ex]{0pt}{4ex} 
$D_{90}$ [Mpc] &  
   77.9 &   63.3 &   43.7 &   24.9 \\ 
\hline
\end{tabularx}
 \begin{tabularx}{\columnwidth}{  c X X X X }
\hline 
\rule{0pt}{4ex} 
 Unmodelled &    SG    & SG     & SG     & SG\\ 
  search   & E & F & G & H\\ 
\hline 
\rule[-2ex]{0pt}{4ex} 
$D_{90}$ [Mpc] &  
    6.8 &    2.3 &    1.2 &    0.5 \\ 
\hline
\end{tabularx}
 \begin{tabularx}{\columnwidth}{ c c c c c c c}
\hline 
\rule{0pt}{4ex} 
 Unmodelled & DS2P    & DS2P     & WNB     & WNB & WNB & WNB\\ 
  search     & A & B & A & B & C & D\\ 
\hline 
\rule[-2ex]{0pt}{4ex} 
$D_{90}$ [Mpc] &  
    0.7 &    0.7 &   66.4 &   71.7 &   15.2 &    9.2 \\ 
\hline
\end{tabularx}
 \begin{tabularx}{\columnwidth}{  c X X X X X}
\hline 
\rule{0pt}{4ex} 
 Unmodelled  & ADI & ADI & ADI & ADI & ADI\\ 
 search       & A   & B   & C   & D   & E\\ 
\rule[-2ex]{0pt}{4ex} 
$D_{90}$ [Mpc] &  
   17.6 &   64.9 &   23.1 &    8.4 &   25.7 \\ 
\hline
\end{tabularx}
\end{center}
\end{table}

The lower rows of Table \ref{tab:table_frb_median_exclusion_distance} show the median of the $D_{90}$ estimates for all other waveforms considered by the generic transient search. We see that SG models spanning central frequencies 70~Hz to 2000~Hz have corresponding median values of $D_{90}$ in the range $78$~Mpc to $0.5$~Mpc; the latter models' performance diminished at higher frequency through detector response. This is also clearly evident for the DS2P ringdown models, which are more likely to encounter a transient burst of noise than SG models due to their longer durations.  Similarly, the median $D_{90}$ values for the higher frequency WNB  models are lower in comparison with the lower frequency models (WNB-A and WNB-B). These median $D_{90}$ values of the 150~Hz and 550~Hz models differ by around a factor of at least 4. Overall, the median $D_{90}$ varies within a range approaching 2 orders of magnitude, reflecting the wide range of models used in the analysis.  

In comparison with $D_{90}$ values obtained in the \ac{O3a} GRB paper \citep{2021ApJ...915...86A} the values in Table \ref{tab:table_frb_median_exclusion_distance} are almost systematically a factor of 2 smaller for the SG and ADI models used in that study. We find that this is a result of the sky locations surveyed by \ac{CHIME} corresponding with a region of weak sensitivity for the Virgo interferometric detector, due to their relative locations on the surface of the Earth. The average antenna responses for the \ac{H1} and \ac{L1} detectors are of order 0.72 and 0.65 respectively; the same metric for the V1 instrument is 0.28. This has a severe effect when V1 is one of only two detectors in a network, a situation that has occurred 55\% of the time for the generic transient analysis of non-repeating FRBs. Looking ahead, this type of sensitivity bias will be a feature of future searches for CHIME/FRB triggers, as well as surveys by other facilities, depending on their location on the Earth.

In Table \ref{tab:table_frb_exclusion} we present the exclusion distances achieved for each of the FRBs analyzed in our joint analysis. For the modelled search we quote values from each of the 3 classes of compact binary progenitor models considered. For the generic transient search we present values of $D_{90}$ for a representative sample of SG, ADI, DS2P and WNB models. We also provide information relating to the times and positions of these events as well as values of the DM, and the inferred 90\% credible intervals on the luminosity distance. Table \ref{tab:table_frb_exclusion} allows comparison of the inferred luminosity distances of each FRB with the $D_{90}$ value for different searches. 

\input{main_exclusion_distance_table}

\begin{figure*}[!t]
  \centering
  \includegraphics[width=\textwidth]{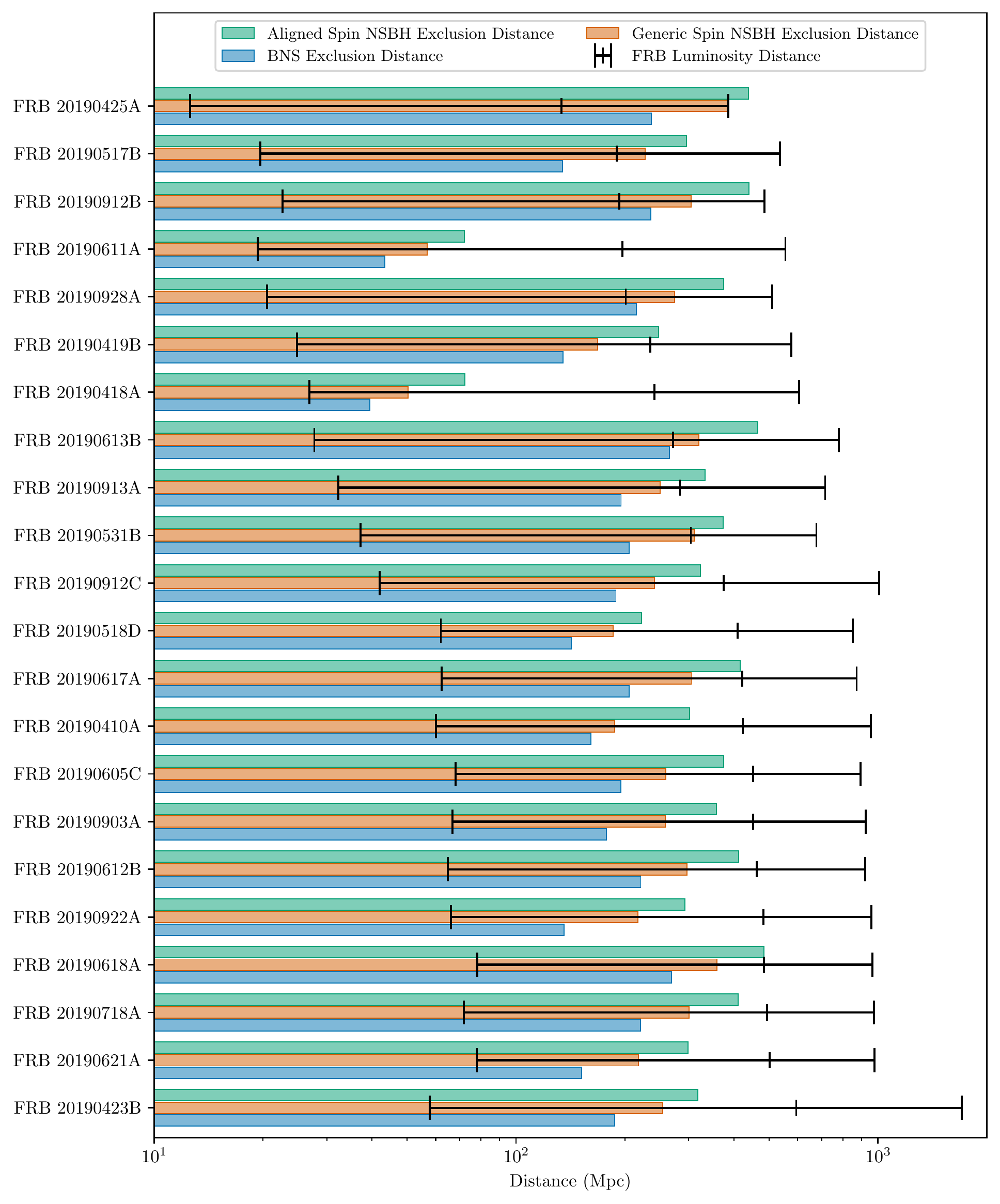}
    \caption{Lower limits on the 90\% confidence level exclusion distances for \ac{BNS} (lower bar), generic spin \ac{NSBH} (middle bar), and aligned spin \ac{NSBH} (upper bar) progenitor systems are shown as found by the modelled search. These are compared to the 90\% credible intervals (whisker plot) on the $D_{L}$ posterior determined by the MCMC method for the FRBs considered in this study.}
  \label{pygrb-ld-exclusion}
\end{figure*}

Fig. \ref{pygrb-ld-exclusion} compares the $D_{90}$ values for the \ac{BNS} and \ac{NSBH} (with generic spin) emission models with the 90\% credible intervals on $D_{L}$ inferred by the MCMC analysis. The plot shows the FRB sample in order of increasing distance. 
No event can be fully excluded from any of the models we have considered for this search, because there is still a sufficient region of space from which the FRB events could have originated that is outside the detection range of the searches performed.

\subsection{RAVEN Analysis Results}

As described in Section \ref{sec:raven_method}, two RAVEN coincidence searches were completed with differing time windows, [$-600$~s,$+120$~s] for the generic transient search and [$-10$~s,$+2$~s] for the modelled search. The generic transient search found 8 coincidences and the modelled search found 1 coincidence. However, none of these were of sufficient significance, as determined by the computed joint false-alarm rate from the two samples, to be distinguished from random coincidences. All of the FRBs in these coincidences had distances that were well beyond the values of $D_{90}$ obtained, with the exception being FRB\,20190518E, a repeat of burst FRB\,20190518A, with 9 episodes occurring during O3a.  Of these 9 repeating episodes, 7 were also analyzed using our generic transient search method, as described earlier. Again, none of the repeating episodes returned a significant false-alarm probability, with the minimum $p$-value across the search of repeating FRB events equal to \lowestRepeaterPValue.

\subsection{Upper Limits on GW Energy}

A measure of the inferred distance to a FRB source also allows one to place constraints on the energy carried in a burst of \ac{GW}s. The GW energy, $E_{\mathrm{GW}}$, emitted by an elliptically polarized \ac{GW} burst signal can be related to the root-sum-square signal amplitude $h_{\mathrm{rss}}$ and the central frequency of the source, $f_{0}$, through \citep{Sutton_2013}:

\begin{equation}\label{eq_E_GW}
    E_{\mathrm{GW}} = \frac{2}{5}\frac{\pi^{2}c^{3}}{G} D_\mathrm{L}^{2}f_{0}^{2} h_{\mathrm{rss}}^{2}\, ,
\end{equation}

\noindent where $D_\mathrm{L}$ is the luminosity distance to the source. As the DMs of FRBs provide a measure of the maximum distance, one can use Eq. (\ref{eq_E_GW}) to place 90\% upper limits on the \ac{GW} energy emitted by each FRB source, $E_{\mathrm{GW}}^{90\%}$. This estimate, calculated using $h_{\mathrm{rss}}^{90\%}$, the 90\% detection upper limit on the root-sum-squared \ac{GW} amplitude, is highly dependent on the detector sensitivity and antenna factors at the time of the FRB as well as the central frequency of the simulated waveform injections.

Table \ref{tab:table_SG_frb_energy} and Table \ref{tab:table_other_frb_energy} provide the upper limits on $E_{\mathrm{GW}}^{90\%}$ for SG models and DS2P or WNB GW burst models respectively.  These limits assume that the FRB distances are at the lower limits of their inferred distance ranges.  Given a large range of models, and since this quantity scales as $h_{\mathrm{rss}}^2~f_{0}^{2}$, one would expect the lower frequency models to provide the most constraining limits. For SG models, the most constraining estimate was \EULSGA for the 70~Hz SG-A model and for the highest frequency model considered, SG-H at 1995~Hz, the upper limit was \EULSGH. These values were obtained for the closest inferred burst in the sample, FRB\,20190425A. The same burst yielded upper limit values in the range \EULRANGEWNB for the WNB model. The DS2P model gave the best constraints, \EULRANGEDSTWOP,  for FRB\,20190531B.

\input{table_SG_frb_energy}
\input{table_other_frb_energy}

For completeness, in Table \ref{tab:table_SG_frb_energy_up} and Table \ref{tab:table_other_frb_energy_up}, we also provide less constraining limits on $E_{\mathrm{GW}}^{90\%}$ based on the upper credible intervals on the distance of each FRB.

\input{table_SG_frb_energy_up}
\input{table_other_frb_energy_up}

Table \ref{tab:repeaters} lists the repeating bursts that were analyzed in the generic transient search.
The most sensitive counterpart to a repeating FRB was for CHIME/FRB event FRB20190825A. The SG injection centered at 1600~Hz (which most closely models an f-mode) was recovered $90\%$ of the time at $h_{\mathrm{\mathrm{rss}}}=2.62\times~10^{-22}$. The distance to this event is $148.1$~Mpc to $149.9$~Mpc. This corresponds to an energy upper limit range of $5.83\,\times\,10^{55}$~erg to $5.98\,\times\,10^{55}$~erg. 

\begin{table*}
    \caption{\label{tab:repeaters} Details of the 3 repeating FRBs analyzed in the generic transient search and their various repeating episodes. The TNS name is provided in the first column. The Network column lists the GW detector network used: H1 = LIGO Hanford, L1 = LIGO Livingston, V1 = Virgo. The total \ac{DM} for each FRB is listed in the DM column and the 90\% credible intervals on the luminosity distance are provided in columns $D_{L}$-low and $D_{L}$-High. 11 total events were analyzed for the three different FRB repeaters considered. For FRB\,20190518A and its associated repeats,  we list only the distance of \citet{Marcote_2020Natur} obtained by galaxy localization.}
\begin{center}
\centering
\label{tab:table_repeaters}
\begin{tabular}{lccccccc}
FRB Name &  UTC Time  & R.A. &Dec. & Network & DM               &  $D_{L}$-Low  &$D_{L}$-high\\ 
         &   [s]       &      &     &         &  [pc cm$^{-3}$]  &   [Mpc]       &        [Mpc]\\ 
\hline 
\hline 
FRB20190817A &14:39:52 &$  4^{\mathrm{h}} 21^{\mathrm{m}} 08^{\mathrm{s}}$ &$ 73^{\circ} 47' $ &H1L1V1 &  189.5 &   19.5 &  539.2 \\ 
FRB20190929C &11:58:29 &$  4^{\mathrm{h}} 22^{\mathrm{m}} 25^{\mathrm{s}}$ &$ 73^{\circ} 40' $ &H1L1V1 &  191.6 &   20.8 &  550.1 \\ 
\hline 
FRB20190518A &18:13:33 &$  1^{\mathrm{h}} 58^{\mathrm{m}} 14^{\mathrm{s}}$ &$ 65^{\circ} 46' $ &L1V1 &  350.5 &   148.1 & 149.9 \\ 
FRB20190518E &18:20:57 &$  1^{\mathrm{h}} 57^{\mathrm{m}} 50^{\mathrm{s}}$ &$ 65^{\circ} 43' $ &L1V1 &  350.0 &   148.1 & 149.9 \\
FRB20190519A &17:50:16 &$  1^{\mathrm{h}} 43^{\mathrm{m}} 44^{\mathrm{s}}$ &$ 65^{\circ} 48' $ &H1V1 &  350.0 &   148.1 & 149.9 \\ 
FRB20190519C &18:10:41 &$  1^{\mathrm{h}} 58^{\mathrm{m}} 00^{\mathrm{s}}$ &$ 65^{\circ} 47' $ &H1V1 &  348.8 &   148.1 & 149.9 \\ 
FRB20190809A &12:50:40 &$  1^{\mathrm{h}} 58^{\mathrm{m}} 16^{\mathrm{s}}$ &$ 65^{\circ} 43' $ &H1L1 &  356.2 &   148.1 & 149.9 \\ 
FRB20190825A &11:48:18 &$  1^{\mathrm{h}} 58^{\mathrm{m}} 07^{\mathrm{s}}$ &$ 65^{\circ} 42' $ &H1L1 &  349.6 &   148.1 & 149.9 \\ 
FRB20190825B &11:51:54 &$  1^{\mathrm{h}} 58^{\mathrm{m}} 04^{\mathrm{s}}$ &$ 65^{\circ} 23' $ &H1L1 &  349.9 &   148.1 & 149.9 \\  
\hline 
FRB20190421A &08:00:04 &$ 13^{\mathrm{h}} 51^{\mathrm{m}} 57^{\mathrm{s}}$ &$ 48^{\circ} 10' $ &H1L1V1 &  225.9 &  125.1 & 1260.8 \\ 
FRB20190702B &03:14:36 &$ 13^{\mathrm{h}} 52^{\mathrm{m}} 25^{\mathrm{s}}$ &$ 48^{\circ} 15' $ &L1V1 &  224.4 &  125.8 & 1257.5 \\ 
\hline
\end{tabular}
\end{center}
\end{table*}

These estimates are well above predictions of the \ac{GW} emissions by the \ac{NS}'s fundamental f-mode. For example \citet{CorsiOwen2011PhRvD} have suggested $E_{\mathrm{GW}} \sim 10^{48}-10^{49}$~erg in \ac{GW} energy emitted at around $1-2$ kHz, although predictions in  \citep{Levin_2011MNRAS,ZLK_2012PhRvD} span a much lower range $E_{\mathrm{GW}} \sim 10^{28}-10^{38}$~erg based on studies that suggest lower effective energy conversion to \ac{GW}s.

\section{The M81 repeater FRB\,20200120E}
\label{sec:M81_burst}
A repeater, FRB\,20200120E, which was discovered \mbox{by CHIME/FRB on} 20 Jan 2020, overlaps with \ac{O3b}.  This burst is at 3.6~Mpc, the closest extragalactic FRB so far discovered \citep{2021ApJ_M81_Burst}. 
This event was shown to be conclusively associated with a globular cluster in the M81 galactic system \citep{2021arXiv210511445K} which supports the possibility that it was formed from an evolved stellar population such as a compact binary system. Due to the proximity and significance of this burst, we discuss it in this paper, despite it being discovered after \ac{O3a}.

The burst FRB\,20200120E was shown to repeat at least 4 times. Two of the repeats occurred after \ac{O3b}; another episode, despite being consistent with the localization of the other associated bursts, had no intensity data saved.  Therefore, we discuss here only the initial burst FRB\,20200120E, for which GW data exists. 

At the time of FRB\,20200120E, only \ac{H1} data was available, thus a generic transient search was not conducted. Likewise, since this is a repeating event, it does not pass our criteria for conducting a modelled search. Due to these restrictions, only a RAVEN coincidence search was conducted within a [$-6000, +6000$]~s time window. No coincidences were found with sufficient significance as determined by the coincident false-alarm rate. Given the relative close proximity of this burst, further repeat emissions will be of interest for \ac{GW} follow-up during \ac{O4} \citep{2020LRR....23}.

%% file: main_exclusion_distance_table.tex

\begin{longrotatetable}
\begin{table}
\begin{center}
\centering
\setlength{\tabcolsep}{3pt} 
\renewcommand{\arraystretch}{1} 
\caption{\label{tab:table_frb_exclusion}Details of the FRB sample and the 90\% exclusion distances for
    each of the events considered in this analysis. The TNS name is provided in the first column. The Network column lists the GW detector network used: H1 = LIGO
    Hanford, L1 = LIGO Livingston, V1 = Virgo. The total \ac{DM} for each FRB is listed in the DM column and the 90\% credible intervals on the luminosity distance of each burst are provided in columns $D_{\mathrm{L}}$-Low and $D_{\mathrm{L}}$-High. Where the generic transient search (Section \ref{sec:Xpipeline-search})
    and the modelled search (Section \ref{sec:cbc-search}) used a different
    IFO network, the network used by the generic transient search is shown in parentheses.
    The last 8 columns show the 90\% confidence exclusion distances
    for each FRB ($D_{90}$) for the following emission scenarios:
    BNS, generic and aligned spin NSBH from the modelled search,
    and from the generic transient search, SG-C, SG-F, ADI-A, DS2P-A and WNB-C;
    for the latter 5 types of GW bursts we assume a total radiated energy $E_{\mathrm{GW}}
    = 10^{-2}\,\mathrm{M}_{\odot}\mathrm{c}^{2}$.}
\centering
\begin{tabular}{lcccccccccccccccc}
         &                &         &       &     &         &                  &               &    &\multicolumn{8}{c}{$D_{90}$ [Mpc]}\\
\cmidrule(lr){9-17}\\ 
FRB Name    &UTC Time  & R.A. &Dec. & Network & DM               &  $D_{L}$-Low  &$D_{L}$-High     &BNS  & Generic  & Aligned  &SG  &SG &ADI &DS2P  &WNB\\ 
                    &             &      &     &         &  [pc cm$^{-3}$]    &   [Mpc]       &        [Mpc]    &     &  NSBH    &   NSBH &C   &F &A &A  &C\\ 
\hline 
FRB\,20190410A &12:19:41 &$ 17^{\mathrm{h}} 33^{\mathrm{m}} 43^{\mathrm{s}}$ &$ -2^{\circ} 10' $ &L1V1 &  267.8 &   60.1 &  956.6 &  161.1 &  187.5 &  301.4 &   36.3 &    1.1 &   14.8 &    0.6 &    6.4 &\\ 
FRB\,20190418A &22:34:17 &$  4^{\mathrm{h}} 21^{\mathrm{m}} 07^{\mathrm{s}}$ &$ 15^{\circ} 27' $ &V1 &  184.5 &   26.9 &  605.3 &   39.5 &   50.4 &   72.3 &-  &-  &-  &-  &-  &\\ 
FRB\,20190419B &22:38:24 &$ 17^{\mathrm{h}} 02^{\mathrm{m}} 02^{\mathrm{s}}$ &$ 86^{\circ} 44' $ &L1V1 &  165.2 &   24.8 &  575.7 &  135.0 &  168.2 &  247.6 &   33.7 &    1.1 &   10.2 &    0.5 &    6.2 &\\ 
FRB\,20190423B &13:51:43 &$ 19^{\mathrm{h}} 54^{\mathrm{m}} 44^{\mathrm{s}}$ &$ 26^{\circ} 19' $ &H1V1 &  585.0 &   57.8 & 1704.6 &  187.1 &  254.5 &  318.0 &   12.6 &    0.3 &    5.6 &    0.2 &    3.1 &\\ 
FRB\,20190425A &10:47:49 &$ 17^{\mathrm{h}} 02^{\mathrm{m}} 47^{\mathrm{s}}$ &$ 21^{\circ} 30' $ &H1L1V1 &  128.1 &   12.6 &  385.9 &  236.3 &  386.8 &  437.9 &   66.1 &    3.2 &   27.2 &    0.1 &   20.6 &\\ 
FRB\,20190517B &20:33:37 &$  4^{\mathrm{h}} 16^{\mathrm{m}} 49^{\mathrm{s}}$ &$ 73^{\circ} 10' $ &V1 &  191.4 &   19.6 &  536.3 &  134.5 &  227.7 &  295.7 &-  &-  &-  &-  &-  &\\ 
FRB\,20190517C &22:06:34 &$  5^{\mathrm{h}} 50^{\mathrm{m}} 57^{\mathrm{s}}$ &$ 26^{\circ} 34' $ &L1V1 &  335.5 &   44.3 & 1030.5 &-  &-  &-  &   40.2 &    1.3 &   10.4 &    0.7 &    7.4 &\\ 
FRB\,20190518D &09:04:35 &$ 12^{\mathrm{h}} 06^{\mathrm{m}} 50^{\mathrm{s}}$ &$ 89^{\circ} 25' $ &H1L1 &  202.5 &   62.0 &  852.0 &  141.9 &  185.4 &  222.1 &   54.3 &    3.2 &   20.9 &    1.0 &   15.9 &\\ 
FRB\,20190531B &08:47:40 &$ 17^{\mathrm{h}} 31^{\mathrm{m}} 26^{\mathrm{s}}$ &$ 49^{\circ} 18' $ &L1V1 &  167.9 &   37.2 &  675.5 &  205.1 &  311.7 &  372.9 &   55.8 &    3.5 &   22.8 &    2.1 &   19.7 &\\ 
FRB\,20190601C &21:13:28 &$  5^{\mathrm{h}} 55^{\mathrm{m}} 06^{\mathrm{s}}$ &$ 28^{\circ} 28' $ &H1L1V1 &  424.1 &  198.7 & 1736.9 &-  &-  &-  &   65.9 &    3.2 &   21.1 &    1.1 &   21.2 &\\ 
FRB\,20190604G &23:12:19 &$  8^{\mathrm{h}} 03^{\mathrm{m}} 13^{\mathrm{s}}$ &$ 59^{\circ} 32' $ &L1V1 &  233.0 &   97.1 & 1143.0 &-  &-  &-  &   13.6 &    0.5 &    8.8 &    0.3 &    1.6 &\\ 
FRB\,20190605C &02:20:41 &$ 11^{\mathrm{h}} 14^{\mathrm{m}} 04^{\mathrm{s}}$ &$ -5^{\circ} 18' $ &H1L1V1 (L1V1) &  187.2 &   68.2 &  893.7 &  194.9 &  259.1 &  374.5 &   28.7 &    0.9 &   14.9 &    0.6 &    5.4 &\\ 
FRB\,20190606B &22:19:30 &$  7^{\mathrm{h}} 14^{\mathrm{m}} 42^{\mathrm{s}}$ &$ 86^{\circ} 58' $ &H1L1V1 &  278.0 &  168.6 & 1465.7 &-  &-  &-  &   43.7 &    2.3 &   17.6 &    0.9 &   15.2 &\\ 
FRB\,20190611A &18:52:42 &$  4^{\mathrm{h}} 05^{\mathrm{m}} 12^{\mathrm{s}}$ &$ 73^{\circ} 37' $ &V1 &  196.2 &   19.4 &  554.6 &   43.4 &   56.9 &   72.1 &-  &-  &-  &-  &-  &\\ 
FRB\,20190612B &05:30:37 &$ 14^{\mathrm{h}} 48^{\mathrm{m}} 53^{\mathrm{s}}$ &$ 4^{\circ} 21' $ &H1L1 &  187.1 &   64.9 &  922.2 &  221.1 &  297.1 &  412.2 &   70.4 &    3.3 &   26.3 &    1.1 &   20.8 &\\ 
FRB\,20190613B &18:56:15 &$  4^{\mathrm{h}} 23^{\mathrm{m}} 08^{\mathrm{s}}$ &$ 42^{\circ} 37' $ &H1L1V1 &  285.1 &   27.7 &  780.1 &  265.3 &  319.8 &  465.9 &   78.3 &    4.3 &   27.5 &    1.5 &   23.3 &\\ 
FRB\,20190616A &05:56:30 &$ 15^{\mathrm{h}} 34^{\mathrm{m}} 04^{\mathrm{s}}$ &$ 34^{\circ} 21' $ &H1V1 &  212.7 &  107.3 & 1125.8 &-  &-  &-  &   17.2 &    0.6 &    9.0 &    0.3 &    4.3 &\\ 
FRB\,20190617A &02:12:33 &$ 11^{\mathrm{h}} 49^{\mathrm{m}} 13^{\mathrm{s}}$ &$ 83^{\circ} 50' $ &H1L1V1 (H1V1) &  195.8 &   62.2 &  872.9 &  205.3 &  305.1 &  416.7 &   53.9 &    2.9 &   22.7 &    1.4 &   19.4 &\\ 
FRB\,20190618A &11:42:06 &$ 21^{\mathrm{h}} 24^{\mathrm{m}} 28^{\mathrm{s}}$ &$ 25^{\circ} 25' $ &H1L1 &  228.9 &   78.3 &  964.0 &  268.5 &  359.3 &  484.7 &   80.0 &    4.3 &   24.9 &    1.6 &   25.5 &\\ 
FRB\,20190621A &02:21:17 &$ 12^{\mathrm{h}} 06^{\mathrm{m}} 36^{\mathrm{s}}$ &$ 74^{\circ} 43' $ &L1V1 &  199.5 &   78.0 &  978.1 &  152.0 &  218.1 &  299.1 &   15.4 &    0.4 &    2.2 &    0.3 &    2.9 &\\ 
FRB\,20190624B &22:11:00 &$ 20^{\mathrm{h}} 01^{\mathrm{m}} 07^{\mathrm{s}}$ &$ 73^{\circ} 34' $ &H1V1 &  213.9 &   47.0 &  822.5 &-  &-  &-  &   30.5 &    1.3 &    7.5 &    0.5 &    9.2 &\\ 
FRB\,20190710A &22:09:19 &$  9^{\mathrm{h}} 26^{\mathrm{m}} 32^{\mathrm{s}}$ &$ 63^{\circ} 06' $ &H1L1 &  204.0 &   89.5 &  997.6 &-  &-  &-  &   78.1 &    4.3 &   30.9 &    1.9 &   23.4 &\\ 
FRB\,20190713A &02:19:56 &$  1^{\mathrm{h}} 35^{\mathrm{m}} 49^{\mathrm{s}}$ &$ 72^{\circ} 53' $ &H1V1 &  335.9 &  141.1 & 1436.5 &-  &-  &-  &   28.8 &    0.9 &   10.6 &    0.4 &    7.0 &\\ 
FRB\,20190718A &01:11:16 &$ 13^{\mathrm{h}} 04^{\mathrm{m}} 18^{\mathrm{s}}$ &$ 74^{\circ} 14' $ &H1L1 &  199.6 &   71.8 &  973.4 &  220.4 &  300.9 &  410.3 &   57.6 &    3.5 &   24.7 &    1.6 &   20.8 &\\ 
FRB\,20190722A &18:30:18 &$  6^{\mathrm{h}} 35^{\mathrm{m}} 11^{\mathrm{s}}$ &$ 64^{\circ} 17' $ &L1V1 &  252.0 &   97.8 & 1129.9 &-  &-  &-  &   18.4 &    0.7 &   10.9 &    0.4 &    2.1 &\\ 
\hline
\end{tabular}
\end{center}
\end{table}
\end{longrotatetable}

\begin{longrotatetable}
\begin{table}
\begin{center}
\centering
\setlength{\tabcolsep}{3pt} 
\renewcommand{\arraystretch}{1} 
\centering
\begin{tabular}{lcccccccccccccccc}
         &                         &       &     &         &                  &               &    &\multicolumn{8}{c}{$D_{90}$ [Mpc]}\\
\cmidrule(lr){9-17}\\ 
FRB Name   &UTC Time  & R.A. &Dec. & Network & DM               &  $D_{L}$-Low  &$D_{L}$-High     &BNS  & Generic  & Aligned  &SG  &SG &ADI &DS2P  &WNB\\ 
                       &             &      &     &         &  [pc cm$^{-3}$]    &   [Mpc]       &        [Mpc]    &     &  NSBH    &   BHNS  &C   &F &A &A  &C\\ 
\hline
FRB\,20190812A &04:35:08 &$ 17^{\mathrm{h}} 53^{\mathrm{m}} 14^{\mathrm{s}}$ &$ 50^{\circ} 48' $ &H1L1V1 &  254.6 &  186.5 & 1362.0 &-  &-  &-  &   79.3 &    4.1 &   24.2 &    1.5 &   23.5 &\\ 
FRB\,20190903A &12:25:19 &$  3^{\mathrm{h}} 12^{\mathrm{m}} 01^{\mathrm{s}}$ &$ 21^{\circ} 25' $ &L1V1 &  213.2 &   66.8 &  925.4 &  177.7 &  258.8 &  357.4 &   12.6 &    0.3 &    7.3 &    0.3 &    3.1 &\\ 
FRB\,20190912A &00:50:21 &$ 16^{\mathrm{h}} 13^{\mathrm{m}} 58^{\mathrm{s}}$ &$ 22^{\circ} 13' $ &L1V1 &  211.8 &   97.6 & 1090.5 &-  &-  &-  &   14.7 &    0.5 &    9.4 &    0.3 &    3.5 &\\ 
FRB\,20190912B &08:51:31 &$  0^{\mathrm{h}} 15^{\mathrm{m}} 57^{\mathrm{s}}$ &$ 6^{\circ} 12' $ &H1L1 &  129.2 &   22.7 &  485.0 &  235.7 &  304.1 &  440.5 &   73.8 &    3.6 &   26.6 &    1.1 &   21.3 &\\ 
FRB\,20190912C &09:46:46 &$  1^{\mathrm{h}} 13^{\mathrm{m}} 16^{\mathrm{s}}$ &$ 67^{\circ} 08' $ &H1 &  337.8 &   42.1 & 1005.9 &  188.8 &  241.5 &  323.2 &-  &-  &-  &-  &-  &\\ 
FRB\,20190913A &15:11:12 &$  6^{\mathrm{h}} 40^{\mathrm{m}} 02^{\mathrm{s}}$ &$ 39^{\circ} 39' $ &L1 &  225.9 &   32.3 &  714.3 &  195.1 &  249.7 &  332.8 &-  &-  &-  &-  &-  &\\ 
FRB\,20190922A &00:11:04 &$ 16^{\mathrm{h}} 14^{\mathrm{m}} 10^{\mathrm{s}}$ &$ 68^{\circ} 48' $ &H1V1 &  199.3 &   66.2 &  959.6 &  135.7 &  217.2 &  293.1 &   15.8 &    0.5 &    3.1 &    0.2 &    3.4 &\\ 
FRB\,20190928A &21:32:10 &$ 14^{\mathrm{h}} 00^{\mathrm{m}} 25^{\mathrm{s}}$ &$ 80^{\circ} 06' $ &H1L1V1 &  143.2 &   20.5 &  510.3 &  215.2 &  273.7 &  374.9 &   56.6 &    3.0 &   22.0 &    1.1 &   19.4 &\\ 
FRB\,20190929B &13:32:01 &$  6^{\mathrm{h}} 02^{\mathrm{m}} 53^{\mathrm{s}}$ &$ 11^{\circ} 51' $ &H1L1V1 &  377.0 &  149.0 & 1533.6 &-  &-  &-  &   76.9 &    3.9 &   26.3 &    1.7 &   21.8 &\\ 
\hline
\end{tabular}
\end{center}
\end{table}
\end{longrotatetable}

%% file: table_SG_frb_energy.tex

\setlength{\tabcolsep}{3pt} 
\renewcommand{\arraystretch}{1} 
\begin{table*}
\caption{ The upper limits on the energy emitted through GWs in erg for the generic transient search using the SG waveforms described in Table \ref{tab:table_frb_burst_injections}. The distances represent the lower bounds of 90\% credible intervals from the MCMC inference described in Section \ref{sec:frb_sample}.}
\begin{center}
\centering
\label{tab:table_SG_frb_energy}
\begin{tabular}{l|c|cccccccc}
\hline
\hline
\hspace{6mm}FRB & $D_{\mathrm{L}}$  & SG &SG & SG  & SG  &SG  & SG & SG  & SG\\ 
  &  [Mpc] & A &B & C & D &E & F & G  & H\\ 
\hline 
FRB\,20190410A &   60.1 &$1.5\times10^{52}$&$2.8\times10^{52}$&$4.9\times10^{52}$&$4.1\times10^{53}$&$5.5\times10^{54}$&$5.4\times10^{55}$&$3.0\times10^{56}$&$1.1\times10^{57}$\\ 
FRB\,20190419B &   24.8 &$2.6\times10^{51}$&$4.3\times10^{51}$&$9.7\times10^{51}$&$5.9\times10^{52}$&$9.4\times10^{53}$&$8.9\times10^{54}$&$5.0\times10^{55}$&$1.5\times10^{58}$\\ 
FRB\,20190423B &   57.8 &$5.9\times10^{52}$&$8.9\times10^{52}$&$3.7\times10^{53}$&$3.7\times10^{54}$&$4.6\times10^{55}$&$5.6\times10^{56}$&$3.4\times10^{57}$&$1.1\times10^{58}$\\ 
FRB\,20190425A &   12.6 &$2.5\times10^{50}$&$3.5\times10^{50}$&$6.5\times10^{50}$&$3.4\times10^{51}$&$2.6\times10^{52}$&$2.7\times10^{53}$&$1.6\times10^{54}$&$7.9\times10^{54}$\\ 
FRB\,20190517C &   44.3 &$5.8\times10^{51}$&$8.8\times10^{51}$&$2.2\times10^{52}$&$1.3\times10^{53}$&$2.3\times10^{54}$&$2.1\times10^{55}$&$9.8\times10^{55}$&$3.5\times10^{56}$\\ 
FRB\,20190518D &   62.0 &$9.5\times10^{51}$&$1.3\times10^{52}$&$2.3\times10^{52}$&$9.5\times10^{52}$&$1.1\times10^{54}$&$6.8\times10^{54}$&$3.6\times10^{55}$&$2.0\times10^{56}$\\ 
FRB\,20190531B &   37.2 &$3.2\times10^{51}$&$3.4\times10^{51}$&$7.9\times10^{51}$&$3.3\times10^{52}$&$2.5\times10^{53}$&$2.0\times10^{54}$&$8.1\times10^{54}$&$3.1\times10^{55}$\\ 
FRB\,20190601C &  198.7 &$8.6\times10^{52}$&$1.1\times10^{53}$&$1.6\times10^{53}$&$6.3\times10^{53}$&$1.1\times10^{55}$&$6.8\times10^{55}$&$4.8\times10^{56}$&$1.5\times10^{57}$\\ 
FRB\,20190604G &   97.1 &$1.1\times10^{53}$&$3.2\times10^{53}$&$9.0\times10^{53}$&$3.7\times10^{54}$&$8.7\times10^{55}$&$7.5\times10^{56}$&$3.2\times10^{57}$&$1.2\times10^{58}$\\ 
FRB\,20190605C &   68.2 &$3.0\times10^{52}$&$2.8\times10^{52}$&$1.0\times10^{53}$&$5.2\times10^{53}$&$8.7\times10^{54}$&$9.4\times10^{55}$&$5.2\times10^{56}$&$1.6\times10^{57}$\\ 
FRB\,20190606B &  168.6 &$1.7\times10^{53}$&$1.3\times10^{53}$&$2.7\times10^{53}$&$8.2\times10^{53}$&$1.1\times10^{55}$&$9.6\times10^{55}$&$3.6\times10^{56}$&$1.4\times10^{57}$\\ 
FRB\,20190612B &   64.9 &$8.2\times10^{51}$&$8.5\times10^{51}$&$1.5\times10^{52}$&$7.3\times10^{52}$&$8.0\times10^{53}$&$7.0\times10^{54}$&$3.7\times10^{55}$&$3.6\times10^{56}$\\ 
FRB\,20190613B &   27.7 &$1.2\times10^{51}$&$1.0\times10^{51}$&$2.2\times10^{51}$&$1.3\times10^{52}$&$9.3\times10^{52}$&$7.4\times10^{53}$&$4.2\times10^{54}$&$1.8\times10^{55}$\\ 
FRB\,20190616A &  107.3 &$1.9\times10^{53}$&$2.1\times10^{53}$&$6.9\times10^{53}$&$3.1\times10^{54}$&$3.5\times10^{55}$&$5.1\times10^{56}$&$2.8\times10^{57}$&$8.2\times10^{57}$\\ 
FRB\,20190617A &   62.2 &$9.5\times10^{51}$&$1.3\times10^{52}$&$2.4\times10^{52}$&$9.2\times10^{52}$&$9.2\times10^{53}$&$8.3\times10^{54}$&$4.2\times10^{55}$&$8.8\times10^{55}$\\ 
FRB\,20190618A &   78.3 &$6.0\times10^{51}$&$7.7\times10^{51}$&$1.7\times10^{52}$&$7.0\times10^{52}$&$6.8\times10^{53}$&$5.9\times10^{54}$&$3.0\times10^{55}$&$1.4\times10^{56}$\\ 
FRB\,20190621A &   78.0 &$1.1\times10^{53}$&$1.2\times10^{53}$&$4.6\times10^{53}$&$1.5\times10^{54}$&$5.4\times10^{55}$&$6.5\times10^{56}$&$1.7\times10^{57}$&$4.9\times10^{57}$\\ 
FRB\,20190624B &   47.0 &$1.3\times10^{52}$&$1.9\times10^{52}$&$4.2\times10^{52}$&$1.7\times10^{53}$&$2.9\times10^{54}$&$2.3\times10^{55}$&$1.5\times10^{56}$&$8.3\times10^{56}$\\ 
FRB\,20190710A &   89.5 &$1.1\times10^{52}$&$1.6\times10^{52}$&$2.3\times10^{52}$&$1.0\times10^{53}$&$9.4\times10^{53}$&$7.6\times10^{54}$&$3.3\times10^{55}$&$1.4\times10^{56}$\\ 
FRB\,20190713A &  141.1 &$1.2\times10^{53}$&$1.6\times10^{53}$&$4.3\times10^{53}$&$2.3\times10^{54}$&$4.2\times10^{55}$&$4.4\times10^{56}$&$2.2\times10^{57}$&$6.7\times10^{57}$\\ 
FRB\,20190718A &   71.8 &$1.1\times10^{52}$&$1.1\times10^{52}$&$2.8\times10^{52}$&$1.1\times10^{53}$&$1.1\times10^{54}$&$7.7\times10^{54}$&$3.1\times10^{55}$&$1.2\times10^{56}$\\ 
FRB\,20190722A &   97.8 &$7.0\times10^{52}$&$1.3\times10^{53}$&$5.0\times10^{53}$&$3.3\times10^{54}$&$5.4\times10^{55}$&$4.0\times10^{56}$&$1.6\times10^{57}$&$9.6\times10^{57}$\\ 
FRB\,20190812A &  186.5 &$3.7\times10^{52}$&$4.1\times10^{52}$&$9.9\times10^{52}$&$4.3\times10^{53}$&$4.3\times10^{54}$&$3.7\times10^{55}$&$1.6\times10^{56}$&$5.8\times10^{56}$\\ 
FRB\,20190903A &   66.8 &$9.0\times10^{52}$&$9.8\times10^{52}$&$5.0\times10^{53}$&$4.4\times10^{54}$&$5.5\times10^{55}$&$7.4\times10^{56}$&$3.4\times10^{57}$&$9.2\times10^{57}$\\ 
FRB\,20190912A &   97.6 &$1.2\times10^{53}$&$2.0\times10^{53}$&$7.9\times10^{53}$&$4.6\times10^{54}$&$1.0\times10^{56}$&$8.1\times10^{56}$&$3.8\times10^{57}$&$1.7\times10^{58}$\\ 
FRB\,20190912B &   22.7 &$7.1\times10^{50}$&$9.1\times10^{50}$&$1.7\times10^{51}$&$8.1\times10^{51}$&$6.9\times10^{52}$&$7.1\times10^{53}$&$3.9\times10^{54}$&$1.5\times10^{55}$\\ 
FRB\,20190922A &   66.2 &$5.1\times10^{52}$&$7.7\times10^{52}$&$3.1\times10^{53}$&$1.5\times10^{54}$&$2.4\times10^{55}$&$2.8\times10^{56}$&$1.5\times10^{57}$&$4.7\times10^{57}$\\ 
FRB\,20190928A &   20.5 &$9.9\times10^{50}$&$1.1\times10^{51}$&$2.3\times10^{51}$&$9.2\times10^{51}$&$1.1\times10^{53}$&$8.2\times10^{53}$&$3.7\times10^{54}$&$1.4\times10^{55}$\\ 
FRB\,20190929B &  149.0 &$2.9\times10^{52}$&$3.9\times10^{52}$&$6.7\times10^{52}$&$3.4\times10^{53}$&$2.8\times10^{54}$&$2.6\times10^{55}$&$1.2\times10^{56}$&$4.0\times10^{56}$\\ 
\hline
\end{tabular}
\end{center}
\end{table*}

%% file: table_other_frb_energy.tex

\begin{table*}[htb]
\caption{ The upper limits on the energy emitted through GWs in erg for the generic transient search using the DS2P and WNB waveforms described in Table \ref{tab:table_frb_burst_injections}. The distances represent the lower bounds of 90\% credible intervals from the MCMC inference described in Section \ref{sec:frb_sample}.}
\begin{center}
\centering
\label{tab:table_other_frb_energy}
\begin{tabular}{l|c|cc|cccc}
\hline
\hline
\hspace{6mm}FRB & $D_{\mathrm{L}}$    &DS2P   &DS2P   &WNB   &WNB   &WNB  &WNB\\
  &  [Mpc]     &A  &B   &A   &B  &C &D\\
\hline
FRB\,20190410A &   60.1 &$2.0\times10^{56}$&$1.8\times10^{56}$&$1.2\times10^{53}$&$9.5\times10^{52}$&$3.4\times10^{54}$&$1.0\times10^{55}$\\ 
FRB\,20190419B &   24.8 &$4.4\times10^{55}$&$3.0\times10^{55}$&$1.6\times10^{54}$&$1.8\times10^{52}$&$6.0\times10^{53}$&$1.7\times10^{54}$\\ 
FRB\,20190423B &   57.8 &$2.4\times10^{57}$&$2.6\times10^{57}$&$2.8\times10^{53}$&$3.0\times10^{53}$&$1.4\times10^{55}$&$3.6\times10^{55}$\\ 
FRB\,20190425A &   12.6 &$1.7\times10^{56}$&$4.6\times10^{54}$&$4.8\times10^{50}$&$7.9\times10^{50}$&$1.4\times10^{52}$&$4.7\times10^{52}$\\ 
FRB\,20190517C &   44.3 &$6.7\times10^{55}$&$5.8\times10^{55}$&$2.4\times10^{52}$&$3.1\times10^{52}$&$1.4\times10^{54}$&$7.3\times10^{54}$\\ 
FRB\,20190518D &   62.0 &$6.7\times10^{55}$&$1.1\times10^{56}$&$2.0\times10^{52}$&$2.6\times10^{52}$&$5.8\times10^{53}$&$1.7\times10^{54}$\\ 
FRB\,20190531B &   37.2 &$5.8\times10^{54}$&$6.4\times10^{54}$&$5.7\times10^{51}$&$8.6\times10^{51}$&$1.4\times10^{53}$&$5.6\times10^{53}$\\ 
FRB\,20190601C &  198.7 &$5.5\times10^{56}$&$8.3\times10^{56}$&$1.2\times10^{53}$&$1.6\times10^{53}$&$3.4\times10^{54}$&$8.6\times10^{54}$\\ 
FRB\,20190604G &   97.1 &$1.9\times10^{57}$&$1.6\times10^{57}$&$ - $&$4.9\times10^{54}$&$1.5\times10^{56}$&$3.4\times10^{56}$\\ 
FRB\,20190605C &   68.2 &$2.4\times10^{56}$&$1.7\times10^{56}$&$3.5\times10^{53}$&$1.6\times10^{53}$&$6.2\times10^{54}$&$1.8\times10^{55}$\\ 
FRB\,20190606B &  168.6 &$5.7\times10^{56}$&$9.9\times10^{56}$&$3.6\times10^{53}$&$2.0\times10^{53}$&$4.7\times10^{54}$&$1.3\times10^{55}$\\ 
FRB\,20190612B &   64.9 &$6.2\times10^{55}$&$1.1\times10^{58}$&$1.3\times10^{52}$&$2.1\times10^{52}$&$3.7\times10^{53}$&$1.2\times10^{54}$\\ 
FRB\,20190613B &   27.7 &$6.2\times10^{54}$&$1.1\times10^{55}$&$1.6\times10^{51}$&$2.5\times10^{51}$&$5.4\times10^{52}$&$1.6\times10^{53}$\\ 
FRB\,20190616A &  107.3 &$2.2\times10^{57}$&$2.7\times10^{57}$&$1.1\times10^{54}$&$7.3\times10^{53}$&$2.4\times10^{55}$&$1.4\times10^{56}$\\ 
FRB\,20190617A &   62.2 &$3.6\times10^{55}$&$5.1\times10^{55}$&$3.3\times10^{52}$&$2.7\times10^{52}$&$3.9\times10^{53}$&$1.6\times10^{54}$\\ 
FRB\,20190618A &   78.3 &$4.4\times10^{55}$&$7.0\times10^{55}$&$1.0\times10^{52}$&$1.8\times10^{52}$&$3.6\times10^{53}$&$1.2\times10^{54}$\\ 
FRB\,20190621A &   78.0 &$1.1\times10^{57}$&$4.8\times10^{56}$&$ - $&$9.3\times10^{53}$&$2.8\times10^{55}$&$5.9\times10^{55}$\\ 
FRB\,20190624B &   47.0 &$1.9\times10^{56}$&$3.6\times10^{56}$&$2.8\times10^{52}$&$4.4\times10^{52}$&$1.0\times10^{54}$&$3.7\times10^{54}$\\ 
FRB\,20190710A &   89.5 &$3.9\times10^{55}$&$4.3\times10^{55}$&$1.7\times10^{52}$&$2.6\times10^{52}$&$5.6\times10^{53}$&$1.6\times10^{54}$\\ 
FRB\,20190713A &  141.1 &$2.3\times10^{57}$&$3.7\times10^{57}$&$3.1\times10^{53}$&$5.0\times10^{53}$&$1.5\times10^{55}$&$4.4\times10^{55}$\\ 
FRB\,20190718A &   71.8 &$3.7\times10^{55}$&$6.1\times10^{55}$&$1.7\times10^{52}$&$2.3\times10^{52}$&$4.6\times10^{53}$&$1.4\times10^{54}$\\ 
FRB\,20190722A &   97.8 &$1.1\times10^{57}$&$8.0\times10^{56}$&$9.2\times10^{55}$&$2.7\times10^{54}$&$8.2\times10^{55}$&$1.8\times10^{56}$\\ 
FRB\,20190812A &  186.5 &$2.7\times10^{56}$&$5.3\times10^{56}$&$8.2\times10^{52}$&$1.1\times10^{53}$&$2.4\times10^{54}$&$7.1\times10^{54}$\\ 
FRB\,20190903A &   66.8 &$1.1\times10^{57}$&$7.3\times10^{56}$&$5.0\times10^{53}$&$3.6\times10^{53}$&$1.7\times10^{55}$&$4.5\times10^{55}$\\ 
FRB\,20190912A &   97.6 &$2.0\times10^{57}$&$1.5\times10^{57}$&$7.9\times10^{53}$&$6.6\times10^{53}$&$2.9\times10^{55}$&$8.9\times10^{55}$\\ 
FRB\,20190912B &   22.7 &$7.6\times10^{54}$&$1.4\times10^{55}$&$1.4\times10^{51}$&$1.7\times10^{51}$&$4.3\times10^{52}$&$1.2\times10^{53}$\\ 
FRB\,20190922A &   66.2 &$2.2\times10^{57}$&$3.2\times10^{57}$&$1.5\times10^{54}$&$4.2\times10^{53}$&$1.5\times10^{55}$&$3.9\times10^{55}$\\ 
FRB\,20190928A &   20.5 &$6.2\times10^{54}$&$1.1\times10^{55}$&$1.8\times10^{51}$&$2.6\times10^{51}$&$4.3\times10^{52}$&$1.7\times10^{53}$\\ 
FRB\,20190929B &  149.0 &$1.4\times10^{56}$&$3.0\times10^{56}$&$6.6\times10^{52}$&$7.3\times10^{52}$&$1.8\times10^{54}$&$4.7\times10^{54}$\\ 
\hline
\end{tabular}
\end{center}
\end{table*}

%% file: table_SG_frb_energy_up.tex
\begin{table*}
\caption{ As for Table \ref{tab:table_SG_frb_energy} but with distances based on the the upper bounds of 90\% credible intervals on the luminosity distance.}
\begin{center}
\centering
\label{tab:table_SG_frb_energy_up}
\begin{tabular}{l|c|cccccccc}
\hline
\hline
\hspace{6mm}FRB & $D_{\mathrm{L}}$  & SG &SG & SG  & SG  &SG  & SG & SG  & SG\\
  &  [Mpc] & A &B & C & D &E & F & G  & H\\
\hline
FRB\,20190410A &  956.6 &$3.9\times10^{54}$&$7.2\times10^{54}$&$1.2\times10^{55}$&$1.0\times10^{56}$&$1.4\times10^{57}$&$1.4\times10^{58}$&$7.5\times10^{58}$&$2.7\times10^{59}$\\ 
FRB\,20190419B &  575.7 &$1.4\times10^{54}$&$2.3\times10^{54}$&$5.2\times10^{54}$&$3.2\times10^{55}$&$5.1\times10^{56}$&$4.8\times10^{57}$&$2.7\times10^{58}$&$8.0\times10^{60}$\\ 
FRB\,20190423B & 1704.6 &$5.1\times10^{55}$&$7.7\times10^{55}$&$3.2\times10^{56}$&$3.2\times10^{57}$&$4.0\times10^{58}$&$4.9\times10^{59}$&$2.9\times10^{60}$&$9.4\times10^{60}$\\ 
FRB\,20190425A &  385.9 &$2.4\times10^{53}$&$3.3\times10^{53}$&$6.1\times10^{53}$&$3.2\times10^{54}$&$2.4\times10^{55}$&$2.5\times10^{56}$&$1.6\times10^{57}$&$7.5\times10^{57}$\\ 
FRB\,20190517C & 1030.5 &$3.1\times10^{54}$&$4.7\times10^{54}$&$1.2\times10^{55}$&$6.8\times10^{55}$&$1.2\times10^{57}$&$1.1\times10^{58}$&$5.3\times10^{58}$&$1.9\times10^{59}$\\ 
FRB\,20190518D &  852.0 &$1.8\times10^{54}$&$2.4\times10^{54}$&$4.4\times10^{54}$&$1.8\times10^{55}$&$2.0\times10^{56}$&$1.3\times10^{57}$&$6.9\times10^{57}$&$3.8\times10^{58}$\\ 
FRB\,20190531B &  675.5 &$1.0\times10^{54}$&$1.1\times10^{54}$&$2.6\times10^{54}$&$1.1\times10^{55}$&$8.2\times10^{55}$&$6.7\times10^{56}$&$2.7\times10^{57}$&$1.0\times10^{58}$\\ 
FRB\,20190601C & 1736.9 &$6.6\times10^{54}$&$8.3\times10^{54}$&$1.2\times10^{55}$&$4.8\times10^{55}$&$8.2\times10^{56}$&$5.2\times10^{57}$&$3.6\times10^{58}$&$1.1\times10^{59}$\\ 
FRB\,20190604G & 1143.0 &$1.5\times10^{55}$&$4.5\times10^{55}$&$1.3\times10^{56}$&$5.1\times10^{56}$&$1.2\times10^{58}$&$1.0\times10^{59}$&$4.5\times10^{59}$&$1.6\times10^{60}$\\ 
FRB\,20190605C &  893.7 &$5.1\times10^{54}$&$4.9\times10^{54}$&$1.7\times10^{55}$&$8.9\times10^{55}$&$1.5\times10^{57}$&$1.6\times10^{58}$&$8.9\times10^{58}$&$2.7\times10^{59}$\\ 
FRB\,20190606B & 1465.7 &$1.3\times10^{55}$&$9.6\times10^{54}$&$2.0\times10^{55}$&$6.2\times10^{55}$&$8.3\times10^{56}$&$7.3\times10^{57}$&$2.7\times10^{58}$&$1.0\times10^{59}$\\ 
FRB\,20190612B &  922.2 &$1.7\times10^{54}$&$1.7\times10^{54}$&$3.1\times10^{54}$&$1.5\times10^{55}$&$1.6\times10^{56}$&$1.4\times10^{57}$&$7.4\times10^{57}$&$7.3\times10^{58}$\\ 
FRB\,20190613B &  780.1 &$9.7\times10^{53}$&$8.2\times10^{53}$&$1.8\times10^{54}$&$1.0\times10^{55}$&$7.4\times10^{55}$&$5.8\times10^{56}$&$3.4\times10^{57}$&$1.4\times10^{58}$\\ 
FRB\,20190616A & 1125.8 &$2.1\times10^{55}$&$2.4\times10^{55}$&$7.6\times10^{55}$&$3.4\times10^{56}$&$3.9\times10^{57}$&$5.6\times10^{58}$&$3.1\times10^{59}$&$9.0\times10^{59}$\\ 
FRB\,20190617A &  872.9 &$1.9\times10^{54}$&$2.5\times10^{54}$&$4.7\times10^{54}$&$1.8\times10^{55}$&$1.8\times10^{56}$&$1.6\times10^{57}$&$8.2\times10^{57}$&$1.7\times10^{58}$\\ 
FRB\,20190618A &  964.0 &$9.1\times10^{53}$&$1.2\times10^{54}$&$2.6\times10^{54}$&$1.1\times10^{55}$&$1.0\times10^{56}$&$9.0\times10^{56}$&$4.5\times10^{57}$&$2.1\times10^{58}$\\ 
FRB\,20190621A &  978.1 &$1.7\times10^{55}$&$2.0\times10^{55}$&$7.2\times10^{55}$&$2.3\times10^{56}$&$8.4\times10^{57}$&$1.0\times10^{59}$&$2.6\times10^{59}$&$7.7\times10^{59}$\\ 
FRB\,20190624B &  822.5 &$4.0\times10^{54}$&$5.8\times10^{54}$&$1.3\times10^{55}$&$5.1\times10^{55}$&$8.9\times10^{56}$&$7.0\times10^{57}$&$4.6\times10^{58}$&$2.5\times10^{59}$\\ 
FRB\,20190710A &  997.6 &$1.4\times10^{54}$&$2.0\times10^{54}$&$2.9\times10^{54}$&$1.2\times10^{55}$&$1.2\times10^{56}$&$9.5\times10^{56}$&$4.1\times10^{57}$&$1.7\times10^{58}$\\ 
FRB\,20190713A & 1436.5 &$1.2\times10^{55}$&$1.6\times10^{55}$&$4.4\times10^{55}$&$2.4\times10^{56}$&$4.4\times10^{57}$&$4.6\times10^{58}$&$2.2\times10^{59}$&$6.9\times10^{59}$\\ 
FRB\,20190718A &  973.4 &$2.0\times10^{54}$&$2.1\times10^{54}$&$5.1\times10^{54}$&$2.0\times10^{55}$&$1.9\times10^{56}$&$1.4\times10^{57}$&$5.8\times10^{57}$&$2.3\times10^{58}$\\ 
FRB\,20190722A & 1129.9 &$9.4\times10^{54}$&$1.7\times10^{55}$&$6.7\times10^{55}$&$4.4\times10^{56}$&$7.2\times10^{57}$&$5.3\times10^{58}$&$2.2\times10^{59}$&$1.3\times10^{60}$\\ 
FRB\,20190812A & 1362.0 &$2.0\times10^{54}$&$2.2\times10^{54}$&$5.3\times10^{54}$&$2.3\times10^{55}$&$2.3\times10^{56}$&$2.0\times10^{57}$&$8.7\times10^{57}$&$3.1\times10^{58}$\\ 
FRB\,20190903A &  925.4 &$1.7\times10^{55}$&$1.9\times10^{55}$&$9.6\times10^{55}$&$8.4\times10^{56}$&$1.0\times10^{58}$&$1.4\times10^{59}$&$6.5\times10^{59}$&$1.8\times10^{60}$\\ 
FRB\,20190912A & 1090.5 &$1.5\times10^{55}$&$2.5\times10^{55}$&$9.9\times10^{55}$&$5.8\times10^{56}$&$1.2\times10^{58}$&$1.0\times10^{59}$&$4.7\times10^{59}$&$2.2\times10^{60}$\\ 
FRB\,20190912B &  485.0 &$3.2\times10^{53}$&$4.1\times10^{53}$&$7.7\times10^{53}$&$3.7\times10^{54}$&$3.1\times10^{55}$&$3.3\times10^{56}$&$1.8\times10^{57}$&$6.7\times10^{57}$\\ 
FRB\,20190922A &  959.6 &$1.1\times10^{55}$&$1.6\times10^{55}$&$6.6\times10^{55}$&$3.2\times10^{56}$&$5.0\times10^{57}$&$5.9\times10^{58}$&$3.2\times10^{59}$&$9.8\times10^{59}$\\ 
FRB\,20190928A &  510.3 &$6.1\times10^{53}$&$6.9\times10^{53}$&$1.5\times10^{54}$&$5.7\times10^{54}$&$6.6\times10^{55}$&$5.1\times10^{56}$&$2.3\times10^{57}$&$8.4\times10^{57}$\\ 
FRB\,20190929B & 1533.6 &$3.0\times10^{54}$&$4.1\times10^{54}$&$7.1\times10^{54}$&$3.6\times10^{55}$&$3.0\times10^{56}$&$2.8\times10^{57}$&$1.3\times10^{58}$&$4.2\times10^{58}$\\ 
\hline
\end{tabular}
\end{center}
\end{table*}

%% file: table_other_frb_energy_up.tex
\begin{table*}[htb]
\caption{ As for Table \ref{tab:table_other_frb_energy} but with distances based on the the upper bounds of 90\% credible intervals on the luminosity distance.}
\begin{center}
\centering
\label{tab:table_other_frb_energy_up}
\begin{tabular}{l|c|cc|cccc}
\hline
\hline
\hspace{6mm}FRB & $D_{\mathrm{L}}$    &DS2P   &DS2P   &WNB   &WNB   &WNB  &WNB\\
  &  [Mpc]     &A  &B   &A   &B  &C &D\\
\hline
FRB\,20190410A &  956.6 &$5.0\times10^{58}$&$4.5\times10^{58}$&$3.2\times10^{55}$&$2.4\times10^{55}$&$8.6\times10^{56}$&$2.5\times10^{57}$\\ 
FRB\,20190419B &  575.7 &$2.3\times10^{58}$&$1.6\times10^{58}$&$8.4\times10^{56}$&$9.5\times10^{54}$&$3.2\times10^{56}$&$9.0\times10^{56}$\\ 
FRB\,20190423B & 1704.6 &$2.1\times10^{60}$&$2.2\times10^{60}$&$2.4\times10^{56}$&$2.6\times10^{56}$&$1.2\times10^{58}$&$3.2\times10^{58}$\\ 
FRB\,20190425A &  385.9 &$1.6\times10^{59}$&$4.4\times10^{57}$&$4.5\times10^{53}$&$7.4\times10^{53}$&$1.3\times10^{55}$&$4.4\times10^{55}$\\ 
FRB\,20190517C & 1030.5 &$3.6\times10^{58}$&$3.2\times10^{58}$&$1.3\times10^{55}$&$1.7\times10^{55}$&$7.5\times10^{56}$&$4.0\times10^{57}$\\ 
FRB\,20190518D &  852.0 &$1.3\times10^{58}$&$2.1\times10^{58}$&$3.7\times10^{54}$&$4.9\times10^{54}$&$1.1\times10^{56}$&$3.2\times10^{56}$\\ 
FRB\,20190531B &  675.5 &$1.9\times10^{57}$&$2.1\times10^{57}$&$1.9\times10^{54}$&$2.9\times10^{54}$&$4.5\times10^{55}$&$1.8\times10^{56}$\\ 
FRB\,20190601C & 1736.9 &$4.2\times10^{58}$&$6.4\times10^{58}$&$9.4\times10^{54}$&$1.2\times10^{55}$&$2.6\times10^{56}$&$6.6\times10^{56}$\\ 
FRB\,20190604G & 1143.0 &$2.6\times10^{59}$&$2.2\times10^{59}$&$ - $&$6.8\times10^{56}$&$2.0\times10^{58}$&$4.7\times10^{58}$\\ 
FRB\,20190605C &  893.7 &$4.1\times10^{58}$&$2.9\times10^{58}$&$5.9\times10^{55}$&$2.7\times10^{55}$&$1.1\times10^{57}$&$3.0\times10^{57}$\\ 
FRB\,20190606B & 1465.7 &$4.3\times10^{58}$&$7.5\times10^{58}$&$2.7\times10^{55}$&$1.5\times10^{55}$&$3.6\times10^{56}$&$9.7\times10^{56}$\\ 
FRB\,20190612B &  922.2 &$1.3\times10^{58}$&$2.2\times10^{60}$&$2.7\times10^{54}$&$4.2\times10^{54}$&$7.5\times10^{55}$&$2.5\times10^{56}$\\ 
FRB\,20190613B &  780.1 &$4.9\times10^{57}$&$8.9\times10^{57}$&$1.3\times10^{54}$&$2.0\times10^{54}$&$4.3\times10^{55}$&$1.3\times10^{56}$\\ 
FRB\,20190616A & 1125.8 &$2.4\times10^{59}$&$3.0\times10^{59}$&$1.2\times10^{56}$&$8.0\times10^{55}$&$2.6\times10^{57}$&$1.6\times10^{58}$\\ 
FRB\,20190617A &  872.9 &$7.1\times10^{57}$&$1.0\times10^{58}$&$6.4\times10^{54}$&$5.3\times10^{54}$&$7.7\times10^{55}$&$3.2\times10^{56}$\\ 
FRB\,20190618A &  964.0 &$6.7\times10^{57}$&$1.1\times10^{58}$&$1.5\times10^{54}$&$2.7\times10^{54}$&$5.4\times10^{55}$&$1.8\times10^{56}$\\ 
FRB\,20190621A &  978.1 &$1.8\times10^{59}$&$7.6\times10^{58}$&$ - $&$1.5\times10^{56}$&$4.4\times10^{57}$&$9.2\times10^{57}$\\ 
FRB\,20190624B &  822.5 &$5.9\times10^{58}$&$1.1\times10^{59}$&$8.5\times10^{54}$&$1.4\times10^{55}$&$3.1\times10^{56}$&$1.1\times10^{57}$\\ 
FRB\,20190710A &  997.6 &$4.8\times10^{57}$&$5.4\times10^{57}$&$2.1\times10^{54}$&$3.2\times10^{54}$&$6.9\times10^{55}$&$2.0\times10^{56}$\\ 
FRB\,20190713A & 1436.5 &$2.4\times10^{59}$&$3.8\times10^{59}$&$3.2\times10^{55}$&$5.2\times10^{55}$&$1.6\times10^{57}$&$4.5\times10^{57}$\\ 
FRB\,20190718A &  973.4 &$6.7\times10^{57}$&$1.1\times10^{58}$&$3.2\times10^{54}$&$4.2\times10^{54}$&$8.4\times10^{55}$&$2.6\times10^{56}$\\ 
FRB\,20190722A & 1129.9 &$1.5\times10^{59}$&$1.1\times10^{59}$&$1.2\times10^{58}$&$3.6\times10^{56}$&$1.1\times10^{58}$&$2.3\times10^{58}$\\ 
FRB\,20190812A & 1362.0 &$1.4\times10^{58}$&$2.8\times10^{58}$&$4.4\times10^{54}$&$6.1\times10^{54}$&$1.3\times10^{56}$&$3.8\times10^{56}$\\ 
FRB\,20190903A &  925.4 &$2.1\times10^{59}$&$1.4\times10^{59}$&$9.6\times10^{55}$&$7.0\times10^{55}$&$3.3\times10^{57}$&$8.6\times10^{57}$\\ 
FRB\,20190912A & 1090.5 &$2.5\times10^{59}$&$1.8\times10^{59}$&$9.8\times10^{55}$&$8.2\times10^{55}$&$3.7\times10^{57}$&$1.1\times10^{58}$\\ 
FRB\,20190912B &  485.0 &$3.5\times10^{57}$&$6.4\times10^{57}$&$6.5\times10^{53}$&$7.9\times10^{53}$&$2.0\times10^{55}$&$5.3\times10^{55}$\\ 
FRB\,20190922A &  959.6 &$4.7\times10^{59}$&$6.7\times10^{59}$&$3.2\times10^{56}$&$8.8\times10^{55}$&$3.1\times10^{57}$&$8.2\times10^{57}$\\ 
FRB\,20190928A &  510.3 &$3.8\times10^{57}$&$7.1\times10^{57}$&$1.1\times10^{54}$&$1.6\times10^{54}$&$2.6\times10^{55}$&$1.1\times10^{56}$\\ 
FRB\,20190929B & 1533.6 &$1.5\times10^{58}$&$3.1\times10^{58}$&$7.0\times10^{54}$&$7.8\times10^{54}$&$1.9\times10^{56}$&$5.0\times10^{56}$\\ 
\hline
\end{tabular}
\end{center}
\end{table*}

%% file: conclusions.tex
\section{Conclusions}
\label{sec:astrophysical-implications}

We performed a targeted search for \acp{GW} associated with \acp{FRB} detected by the CHIME/FRB project during \ac{O3a}. As the sources of non-repeating FRBs are currently not known, we ran both a modelled search for \ac{BNS} and \ac{NSBH} signals \citep{Harry:2010fr, Williamson:2014wma} and a generic transient search for generic \ac{GW} transient signals \citep{Sutton2010NJPh, 2012PhRvD..86b2003W}.

Our searches found no significant \ac{GW} event candidates in association with the analyzed FRBs.
We set 90\% confidence lower bounds on the distances to FRB progenitors for several different emission models. 
Additionally, we present 90\% credible intervals on the luminosity distance, $D_{L}$, inferred from the DM measurement of each FRB source. 

The $D_{L}$ information can be used to test models based on the simulated injections used for calculating the $D_{90}$ values of each FRB. However, the significant uncertainties in the relative contributions to the total DM for each FRB produce relatively wide credible intervals for the $D_{L}$ posteriors. We find no FRB event can be fully excluded from any of the models we have considered due to some posterior support on $D_{L}$ existing for the FRB outside the detection range of the analyzes performed.

The results however, as illustrated in Fig. \ref{pygrb-ld-exclusion}, indicate that the \ac{GW} network's detection range is advancing into cosmological volumes where FRB emissions are expected. This is encouraging as we look forward to future \ac{GW} searches at higher sensitivity. Furthermore, the redshifts obtained from the ongoing efforts to localize host galaxies (there are currently 18 FRBs with an associated host galaxy (see \url{http://frbhosts.org/}) could significantly improve the chances of constraining progenitor populations \citep{Heintz_2020ApJ,Bhandari_2021}. 
  
The distance estimates for each FRB allowed us to place 90\% upper limits on the \ac{GW} energy emitted by each FRB source, $E_{\mathrm{GW}}^{90\%}$. For each non-repeating FRB analyzed with a generic transient search, we provided limits on $E_{\mathrm{GW}}^{90\%}$ for a range of emission models.
Repeating FRBs were also analyzed to determine 90\% upper limits on the energy emitted through GWs. 
For the most sensitive repeating FRB analysis in our sample we find an energy upper limit range of $5.83\,\times\,10^{54}$~erg to $5.98\,\times\,10^{55}$~erg, well above the predictions for \ac{GW} emissions from the fundamental f-modes of \acp{NS}. Based on Equation 5, an FRB event such as that associated with SGR 1935+2154 occurring during O3a would have allowed the search to probe the more optimistic of these estimates allowing limits, $E_{\mathrm{GW}}\,\sim\,10^{47}$\,erg, assuming a generic burst waveform emitting at roughly $1$~kHz at 10~kpc.

We also analyzed the repeater, FRB\,20200120E, discovered on 20 Jan 2020 during \ac{O3b}. A RAVEN  \citep{urban2016, cho2019low} coincidence search for any previously detected compact binary coalescence GW events was conducted within a [$-6000$, $+6000$]\,s time window around the first burst of this repeater. No coincidences were found with sufficient significance to be distinguished from random coincidences, as determined by the computed joint false-alarm rate from the two samples. 

%% file: acknowledgments.tex
This material is based upon work supported by NSF’s LIGO Laboratory which is a major facility
fully funded by the National Science Foundation.
The authors also gratefully acknowledge the support of
the Science and Technology Facilities Council (STFC) of the
United Kingdom, the Max-Planck-Society (MPS), and the State of
Niedersachsen/Germany for support of the construction of Advanced LIGO 
and construction and operation of the GEO\,600 detector. 
Additional support for Advanced LIGO was provided by the Australian Research Council.
The authors gratefully acknowledge the Italian Istituto Nazionale di Fisica Nucleare (INFN),  
the French Centre National de la Recherche Scientifique (CNRS) and
the Netherlands Organization for Scientific Research (NWO), 
for the construction and operation of the Virgo detector
and the creation and support  of the EGO consortium. 
The authors also gratefully acknowledge research support from these agencies as well as by 
the Council of Scientific and Industrial Research of India, 
the Department of Science and Technology, India,
the Science \& Engineering Research Board (SERB), India,
the Ministry of Human Resource Development, India,
the Spanish Agencia Estatal de Investigaci\'on (AEI),
the Spanish Ministerio de Ciencia e Innovaci\'on and Ministerio de Universidades,
the Conselleria de Fons Europeus, Universitat i Cultura and the Direcci\'o General de Pol\'{\i}tica Universitaria i Recerca del Govern de les Illes Balears,
the Conselleria d'Innovaci\'o, Universitats, Ci\`encia i Societat Digital de la Generalitat Valenciana and
the CERCA Programme Generalitat de Catalunya, Spain,
the National Science Centre of Poland and the European Union – European Regional Development Fund; Foundation for Polish Science (FNP),
the Swiss National Science Foundation (SNSF),
the Russian Foundation for Basic Research, 
the Russian Science Foundation,
the European Commission,
the European Social Funds (ESF),
the European Regional Development Funds (ERDF),
the Royal Society, 
the Scottish Funding Council, 
the Scottish Universities Physics Alliance, 
the Hungarian Scientific Research Fund (OTKA),
the French Lyon Institute of Origins (LIO),
the Belgian Fonds de la Recherche Scientifique (FRS-FNRS), 
Actions de Recherche Concertées (ARC) and
Fonds Wetenschappelijk Onderzoek – Vlaanderen (FWO), Belgium,
the Paris \^{I}le-de-France Region, 
the National Research, Development and Innovation Office Hungary (NKFIH), 
the National Research Foundation of Korea,
the Natural Science and Engineering Research Council Canada,
Canadian Foundation for Innovation (CFI),
the Brazilian Ministry of Science, Technology, and Innovations,
the International Center for Theoretical Physics South American Institute for Fundamental Research (ICTP-SAIFR), 
the Research Grants Council of Hong Kong,
the National Natural Science Foundation of China (NSFC),
the Leverhulme Trust, 
the Research Corporation, 
the Ministry of Science and Technology (MOST), Taiwan,
the United States Department of Energy,
and
the Kavli Foundation.
The authors gratefully acknowledge the support of the NSF, STFC, INFN and CNRS for provision of computational resources.

This work was supported by MEXT, JSPS Leading-edge Research Infrastructure Program, JSPS Grant-in-Aid for Specially Promoted Research 26000005, JSPS Grant-in-Aid for Scientific Research on Innovative Areas 2905: JP17H06358, JP17H06361 and JP17H06364, JSPS Core-to-Core Program A. Advanced Research Networks, JSPS Grant-in-Aid for Scientific Research (S) 17H06133 and 20H05639 , JSPS Grant-in-Aid for Transformative Research Areas (A) 20A203: JP20H05854, the joint research program of the Institute for Cosmic Ray Research, University of Tokyo, National Research Foundation (NRF), Computing Infrastructure Project of KISTI-GSDC, Korea Astronomy and Space Science Institute (KASI), and Ministry of Science and ICT (MSIT) in Korea, Academia Sinica (AS), AS Grid Center (ASGC) and the Ministry of Science and Technology (MoST) in Taiwan under grants including AS-CDA-105-M06, Advanced Technology Center (ATC) of NAOJ, and Mechanical Engineering Center of KEK.

We acknowledge that CHIME is located on the traditional, ancestral, and unceded territory of the Syilx/Okanagan people.

We thank the Dominion Radio Astrophysical Observatory, operated by the National Research Council Canada, for gracious hospitality and expertise. CHIME is funded by a grant from the Canada Foundation for Innovation (CFI) 2012 Leading Edge Fund (Project 31170) and by contributions from the provinces of British Columbia, Qu\'ebec and Ontario. The CHIME/FRB Project is funded by a grant from the CFI 2015 Innovation Fund (Project 33213) and by contributions from the provinces of British Columbia and Qu\'ebec, and by the Dunlap Institute for Astronomy and Astrophysics at the University of Toronto. Additional support was provided by the Canadian Institute for Advanced Research (CIFAR), McGill University and the McGill Space Institute via the Trottier Family Foundation, and the University of British Columbia. The Dunlap Institute is funded through an endowment established by the David Dunlap family and the University of Toronto. Research at Perimeter Institute is supported by the Government of Canada through Industry Canada and by the Province of Ontario through the Ministry of Research \& Innovation. The National Radio Astronomy Observatory is a facility of the National Science Foundation (NSF) operated under cooperative agreement by Associated Universities, Inc. FRB research at UBC is supported by an NSERC Discovery Grant and by the Canadian Institute for Advanced Research. The CHIME/FRB baseband system is funded in part by a CFI John R. Evans Leaders Fund award to IHS.

We would like to thank all of the essential workers who put their health at risk during the COVID-19 pandemic, without whom we would not have been able to complete this work.